\documentstyle[prb,aps,epsfig,graphicx,tabularx, multicol]{revtex}
 
\title{Capacitance of a quantum dot from the  channel-anisotropic two-channel Kondo model}
 
\author{
Karyn Le Hur$^{*}$
and Georg Seelig$^{\dagger}$
}
 
\address{ D\'epartement de Physique Th\'eorique,
 Universit\'e de Gen\`eve,
CH-1211 Gen\`eve 4, Switzerland}
 
\date{\today}

\newcommand{\mybeginwide}{
    \end{multicols}\widetext
    \vspace*{-0.2truein}\noindent
    \hrulefill\hspace*{3.6truein}
}
\newcommand{\myendwide}{
    \hspace*{3.6truein}\noindent\hrulefill
    \begin{multicols}{2}\narrowtext\noindent
}
 
\begin{document}
\maketitle
\bigskip
 
\begin{abstract}

We investigate the charge fluctuations of a large quantum dot coupled to
a two-dimensional electron gas via a quantum point contact following the work of Matveev\cite{m1,m2}. We limit our
 discussion to the case where exactly {\it two} channels enter
the dot and we discuss the role of an anisotropy
between the transmission coefficients (for these two channels) 
at the constriction.
Experimentally, a channel-anisotropy can be introduced  applying
 a relatively weak  in-plane magnetic field to the system when only
 one ``orbital'' channel is open. The magnetic field leads 
to different transmission amplitudes for spin-up and spin-down electrons. 
 In a  strong magnetic field the anisotropic two-channel
 limit corresponds to two (spin-polarized) orbital channels entering the dot.
 The physics of the charge fluctuations 
can be captured using a mapping on 
the channel-anisotropic two-channel Kondo model. 
For the case of weak reflection at the point contact this has already
 briefly been  stressed by one of us in PRB {\bf 64}, 161302R (2001).
This mapping is also appropriate to discuss the conductance behavior of a
two-contact set-up in strong magnetic field. 
 Here, we elaborate on this approach and also discuss an
 alternative solution using  a mapping on a channel-isotropic Kondo model.
In addition we consider the limit of weak transmission. 
  We show that the Coulomb-staircase behavior of the charge in the
dot as a function of the gate voltage, is already smeared out by a
small channel-anisotropy both in the weak- and strong transmission limits. 

\end{abstract}
 
 
\begin{multicols}{2}                            
\narrowtext
 
\section{introduction}

In the past few years a great amount of work has been devoted to studying the Kondo effect in mesoscopic systems\cite{kouwenhoven}. A motivation for these efforts was the recent experimental observation of the Kondo effect in tunneling through a small quantum dot\cite{goldhaber,cronenwett}. In these experiments the effective (or excess) electronic spin of the dot acts as a  magnetic impurity.

\begin{figure}
\includegraphics[scale=0.4]{{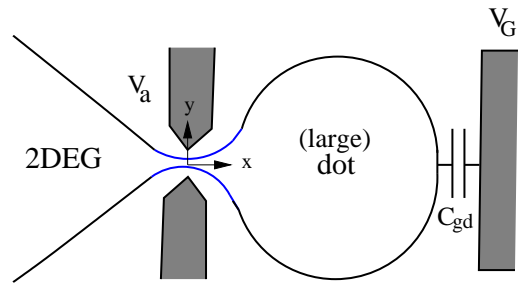}}
\vskip 0.3cm
\caption{\label{dot}
A large quantum dot is coupled to a 2DEG via a quantum point contact. The number of electrons in the dot can be controlled through the gate voltage $V_G$. The auxiliary voltage $V_a$ can be used to open or close the point contact and thus to adjust the reflection amplitudes for the transport channels in the QPC.}
\end{figure}

A different set of problems relating the Kondo effect to the physics of quantum dots is encountered when studying fluctuations of the charge of a large Coulomb-blockaded quantum dot. The setup we have in mind  consists of a large 
quantum dot coupled to a reservoir via a quantum point contact (QPC) and capacitively coupled (with a capacitance $C_{gd}$) to a back-gate (see Fig.~\ref{dot}). The amount of charge on the dot can be changed through the gate voltage $V_G$. The term ``large'' implies that the spacing $\Delta$ of the energy levels on the dot (almost) vanishes and is much smaller than the dot's charging energy $E_C=e^2/(2 C_{gd})$. We assume that the capacitance between the 
two-dimensional electron gas (2DEG) and the dot can be neglected.  
 
 There are two limits in which a mapping to a Kondo Hamiltonian can be used to calculate the charge on the quantum dot. These two limits roughly correspond to the cases of very strong and very weak reflection at the QPC. Before elaborating on this point we  will here give the standard multi-channel Kondo Hamiltonian 
\begin{eqnarray}\label{kondo0}
H_K&=&H_{Kin}+\sum_{\alpha}\big[ J_{\alpha,z}s_{\alpha,z}(0)S^i_{z}\\
&+&J_{\alpha,\perp}\left(s_{\alpha,x}(0)S^i_{x}+s_{\alpha,y}(0)S^i_{y}\right)\big]\nonumber
\end{eqnarray}
in a very general form to be used as a point of reference for later discussions\cite{cox}. Here $\vec{S}$ is an impurity spin and $\vec{s}_\alpha(0)$ is the spin of conduction electrons with flavor $\alpha$ at the place of the impurity. We will consider only the two-channel case $\alpha=1,2$. In the discussion of the Coulomb blockade problem we will encounter Kondo Hamiltonians which are both spin- ($ J_{\alpha,z}\neq J_{\alpha,\perp}$) and channel-anisotropic ($ J_{1,\perp}\neq J_{2,\perp}$).
We now want to return to the original problem and  first want to discuss the limit of weak tunneling between the dot and the reservoir (strong reflection at the QPC). For low enough temperatures ($T\ll E_C$) the charge on the dot is quantized. When the gate voltage $V_G$ is increased the charge on the dot changes in a step-like manner. This behavior is referred to as a Coulomb staircase. Charge fluctuations  are important only for those values of the gate voltage at which two neighboring charge states (e.g. with $n$ and $n+1$ electrons on the dot) are energy degenerate. Matveev demonstrated the equivalence of the effective Hamiltonian describing the charge dynamics close to such a degeneracy point to a spin-anisotropic Kondo Hamiltonian\cite{m2}. If the dot and the reservoir are connected by $m$ transport channels with the same transmission probability the problem can be mapped on a channel-isotropic m-channel Kondo model \cite{n1} with $J_{\alpha,z}=0$ and $J_{\alpha,\perp}\ll 1$\cite{m2}. The similarity of the charge dynamics on the dot to the spin dynamics in a Kondo model was already observed by Glazman and Matveev in there discussion on the conductance of a metallic grain (a quantum dot) tunnel-coupled to two reservoirs\cite{glazman}.

The limit of a point contact with one (electrons without spin) or two  (electrons with  spin) transport channels close to perfect transmission has also been treated. An effectively one-dimensional model for the dot-QPC system, which makes it possible to calculate the total charge on the dot using bosonization, was developed by Flensberg\cite{flensberg} and by Matveev\cite{m1}. Matveev\cite{m1} also calculated  the dot's capacitance as a function of the gate voltage and the reflection amplitude at the point contact. In his calculation he demonstrated that the Hamiltonian of the original problem can be mapped onto a Kondo Hamiltonian in the generalized Toulouse limit ($J_{1,z}=J_{2,z}=2\pi v_F$). In this limit the two-channel Kondo model is exactly solvable as was shown by Emery and Kivelson\cite{emery}. It was one of the main results of Ref.~\onlinecite{m1} that Coulomb blockade oscillations (accompanied by a logarithmic singularity in the expressions for the charge or the capacitance close to ``half-integer'' values of the number of electrons in the dot) persist also in the limit of weak reflection, as recently checked experimentally by Berman et al.\cite{Ashoori}. They are completely smeared out only if at least one transport channel is at perfect transmission, the charge in the dot then increases linearly with the gate voltage \cite{flensberg,m1,van}. A review of these results and more generally of interaction effects in quantum dots can be found in  Ref.~\onlinecite{aleiner}.   

Extending  our previous work\cite{karyn} we here want to discuss the effect of  a {\it channel-anisotropy} on the results for the capacitance in the two limits of almost perfect transmission and almost total reflection discussed above. We will show that the Coulomb staircase behavior is smeared out by a small
channel anisotropy both in the weak and strong reflection limit. We will concentrate on the case of two transport channels. Channel-anisotropy then means that the reflection (or transmission) amplitudes for the two channels are different. Such a situation can be realized applying a magnetic field to the sample. A weak in-plane magnetic field leads to different reflection amplitudes for spin-up and spin-down electrons. In a strong magnetic field electrons are spin polarized and the number of open channels in the QPC and their respective reflection amplitudes depend on the opening of the QPC. The confinement potential at the QPC and thus its opening  can be tuned changing the auxiliary gate voltage $V_a$ (see Fig.~\ref{dot}).  In the limit of small reflection an intuitive physical picture for the transition from the channel symmetric case\cite{m1} to the case of a weak asymmetry between the two channels can be obtained \cite{karyn} from the use of the {\it channel-anisotropic} two-channel Kondo model at the Emery-Kivelson line \cite{fabrizio}. This model has also been 
investigated in Ref.~\onlinecite{Ye}. 
The two coupling constants of the asymmetric Kondo model are  directly related to the reflection amplitudes of the two transport channels in the QPC. In this paper we will elaborate on this mapping and discuss its limitations. Although the physical picture associated with the {\it channel-anisotropic} two-channel  Kondo model is very appealing, the Coulomb blockade problem in the high transparency limit with two transport channels with different reflection coefficients can also be solved through a mapping on the  {\it channel-isotropic} two-channel  Kondo model as we will show. Such a procedure was applied by Matveev and Furusaki\cite{m3} when they calculated the inelastic cotunneling through a quantum dot strongly coupled to two quantum point contacts with one transport channel each (electrons in a strong magnetic field). We will see furthermore that in the case of tunnel-coupling (small transmission) an asymmetry between transmission amplitudes of different channels will lead to a mapping of the Hamiltonian on a channel-anisotropic Kondo model with ($J_{1,z}=J_{2,z}=0)$.

Our paper is structured as follows: In a first part we consider the weak coupling limit. We start by developing the model Hamiltonian in Sec.~\ref{model}. In Sec.~\ref{anisotropic} we discuss the mapping on the channel-anisotropic two-channel Kondo model and use it to calculate the capacitance of the quantum dot in a weak magnetic field. In the following section (Sec.~\ref{isotropic}) we show that the same results can also be found from a mapping to a channel-isotropic Kondo model and in addition look at the case of a strong magnetic field applied to the QPC-dot system. In Sec.~\ref{renorm} a simple scaling argument is used to rederive the main (exact) results of Secs.~\ref{anisotropic} and \ref{isotropic}. In Sec. VI, we will show that the 
channel-anisotropic two-channel Kondo model can also be used to discuss the conductance in a two-contact set-up with a strong  in-plane magnetic field.
The limit of small transmission through the quantum dot will finally be considered in   Sec.~\ref{sec_t_small}.

\section{The Model}\label{model}

We here consider a large quantum dot weakly coupled to a 2DEG via a  point 
contact\cite{m1} (see Fig.~\ref{dot}).  The shape of the QPC is defined trough metallic gates put on top of a 2DEG. The lateral confinement potential in the QPC can be controlled changing the voltage applied to these gates. As a consequence of the lateral confinement the conductance of the QPC is quantized  and electron motion in 
the vicinity of the QPC  is essentially one-dimensional. 
The one-dimensional wave-function $\Psi^n_\sigma$ for motion along the $x$ 
axis is characterized by the spin index $\sigma=\uparrow,\downarrow$ and 
the (orbital) channel quantum number $n$ due to the lateral confinement \cite{Markus1}. Note, that from here on we will  use the word channel to mean electrons with a certain pair of indices $\sigma ,n$. As a short notation we introduce the channel index $\alpha=\sigma,n$ and we denote the wave-function of an electron in this channel by $\Psi_\alpha$. In the further analysis we neglect transport channels that are totally reflected at the QPC, since electrons from these channels are confined to the reservoir and will not contribute to the charging of the dot. We will concentrate on the case where only two channels are open. This can be realized in two different ways. If a weak (or zero) in-plane magnetic field is applied (see Sec.~\ref{anisotropic}), the two channels correspond to the two spin polarizations of electrons in the lowest ($n=1$) energy eigenstate of the lateral Hamiltonian ($\Psi_1=\Psi^1_\uparrow,\Psi_2=\Psi^1_\downarrow$). The role of the magnetic field is to introduce an asymmetry between the reflection coefficients for spin-up and spin-down electrons \cite{karyn}. In a strong magnetic field the electrons are spin-polarized and the two-channel case corresponds to two orbital channels entering the constriction (e.g. $\Psi_1=\Psi^1_\uparrow,\Psi_2=\Psi^2_\uparrow$); See Sec.~\ref{isotropic}.

Since we are interested in energies much smaller than the Fermi energy we linearize the spectrum around the Fermi points. Furthermore it is convenient to decompose the wave-function into a right-going and a left-going contribution $\Psi_\alpha={\rm exp}(ik_Fx)\Psi_{\alpha,R}(x)+{\rm exp}(-ik_Fx)\Psi_{\alpha,L}(x)$.  For two open transport channels the Hamiltonian for the electrons in the point contact is
\begin{eqnarray}\label{kinetic1}
H_{Kin} =-iv_F\sum_{\alpha=1,2}\int^{\infty}_{-\infty}&dx&\left[\Psi^{\dagger}_{\alpha,R}(x)\partial_x\Psi_{\alpha,R}(x)\right.\\
&-&\left.\Psi^{\dagger}_{\alpha,L}(x)\partial_x\Psi_{\alpha,L}(x)\right],\nonumber
\end{eqnarray}
where the sum is over (open) quantum channels. Electron interactions in the quantum dot are taken into account via the Coulomb Hamiltonian
\begin{equation}\label{coulomb1}
H_C=E_C(Q-N)^2-E_CN^2.
\end{equation}
The parameter $eN=V_GC_{gd}$ is  proportional to the gate voltage and $E_C=e^2/(2C_{gd})$ is the charging energy. In addition we allow for a weak backscattering\cite{kane} in the point contact: 
\begin{equation}\label{bs1}
H_{bs}=v_F\sum_{\alpha=1,2}|r_\alpha|\left[\Psi^{\dagger}_{\alpha,R}(0)\Psi_{\alpha,L}(0)+h.c.\right].
\end{equation}
The transmission is supposed to be globally adiabatic \cite{GLK}.
The reflection amplitudes $|r_1|$ and $|r_2|$ can be tuned applying a voltage to the gates defining the QPC. The main goal in the following  will be to calculate  the shift in the ground state energy $\delta \epsilon$ due to the backscattering in the point contact. From the correction $\delta \epsilon$ it is possible to obtain  the average charge in the dot and to dot's capacitance via
\begin{eqnarray}\label{fundamental}
\langle Q \rangle &=& C_{gd}V_G-\frac{\partial(\delta \epsilon)}{\partial V_{G}},\\
C&=&\frac{\partial \langle Q \rangle}{\partial V_{G}}=C_{gd}-\frac{\partial^2(\delta \epsilon)}{\partial {V_G}^2}.\nonumber
\end{eqnarray}

To manipulate the Hamiltonian of the one-dimensional interacting system we have introduced above, it is convenient to use the bosonization technique. Bosonizing in the standard way we express the Fermi field operators through the bosonic fields $\phi_\alpha(x)$ and $\theta_\alpha(x)$ and write \cite{Haldane,MG}
\begin{equation}\label{boso}
\Psi_{\alpha,R/L}(x)=\frac{1}{\sqrt{2\pi a}}{\rm e}^{i\sqrt{\pi}\left(-p\phi_\alpha(x)+\theta_\alpha(x)\right)}.
\end{equation}
Here $p=1$ for right-movers ($R$) and $p=-1$ for left-movers ($L$). The bosonic fields obey the commutation relations
\begin{eqnarray}
\left[\phi_\alpha(x),\theta_\beta(y)\right]&=&\frac{i}{2}{\rm sgn}(x-y)\delta_{\alpha ,\beta},\\
\left[\phi_\alpha(x),\partial_y\theta_\beta(y)\right]&=&i\delta (x-y)\delta_{\alpha ,\beta},\nonumber\\
\left[\phi_\alpha(x),\phi_\beta(y)\right]&=&\left[\theta_\alpha(x),\theta_\beta(y)\right]=0.\nonumber
\end{eqnarray}
 In the new variables the kinetic energy takes the form
\begin{equation}\label{kinetic2}
H_{Kin}=v_F\sum_{\alpha=1,2}\int^{\infty}_{-\infty}dx\left[\left(\partial_x\phi_\alpha \right)^2+\left(\partial_x\theta_\alpha\right)^2\right].
\end{equation}
To bosonize the Coulomb Hamiltonian we use the relation $:\Psi^\dagger_{\alpha,L}\Psi_{\alpha,L}+\Psi^\dagger_{\alpha,R}\Psi_{\alpha,R}:=-\partial_x \phi_\alpha/\sqrt{\pi}$. The total charge in the dot then is 
\begin{equation}
Q=\sum_{\alpha=1,2}\int^\infty_0dx\left(:\Psi^\dagger_{\alpha,R}\Psi_{\alpha,R}+\Psi^\dagger_{\alpha,L}\Psi_{\alpha,L}:\right).
\end{equation}
The charge in the dot is now measured in units of $e$. We neglect both finite
size effects in the dot \cite{aleiner,aleiner2} (supposed to be ideal with no dephasing process) 
and mesoscopic corrections to the capacitance $C_{gd}$ \cite{MarPret,Mar2}.
Expressing the charge in terms of the bosonic variables we obtain $Q=\left(\phi_1(0)+\phi_2(0)\right)/\sqrt{\pi}$. The term at spatial infinity is independent of the gate voltage; We choose $\phi_i(\infty)=0$ in such a way that the
total charge Q on the dot is zero when $N=0$. Consequently we get
\begin{equation}\label{coulomb2}
H_C=\frac{E_C}{\pi}\left(\sum_{\alpha=1,2}\phi_\alpha (0)-\sqrt{\pi}N\right)^2
\end{equation}
for the Coulomb Hamiltonian. 
In writing Eq.~(\ref{coulomb2}) we have only considered the $Q$-dependent part
of Eq.~(\ref{coulomb1}).
Finally we also have to bosonize the backscattering Hamiltonian which leads us to 
\begin{equation}\label{bs2}
H_{bs}=\frac{v_F}{\pi a}\sum_{\alpha=1,2}|r_\alpha|\cos\left(\sqrt{4\pi}\phi_\alpha(0)\right).
\end{equation}
Now we introduce the new standard variables $\phi_{c,s}=\frac{1}{\sqrt{2}}\left(\phi_1(x)\pm\phi_2(x)\right)$ and $\theta_{c,s}=\frac{1}{\sqrt{2}}
\left(\theta_1(x)\pm\theta_2(x)\right)$ where the positive sign belongs to the label $c$ (charge) and the negative sign belongs to $s$ (spin). Note that in the case of a strong magnetic field electrons are spin-polarized. The index $s$ then labels a ``pseudospin'' and not physical spin states.  The kinetic energy in the new variables has  the 
standard form of Eq.~(\ref{kinetic2}). The Coulomb Hamiltonian 
\begin{equation}\label{coulomb3}
H_C=\frac{2E_C}{\pi}\left(\phi_c(0)-\sqrt{\pi/2}N\right)^2
\end{equation}
is a function of the charge field only which is the main motivation for the introduction of this new set of variables. Furthermore we get for the backscattering part
\begin{eqnarray}\label{bs3}
H_{bs}&=&\frac{v_F}{\pi a}(|r_1|+|r_2|)\cos\left(\sqrt{2\pi}\phi_c(0)\right)\cos\left(\sqrt{2\pi}\phi_s(0)\right)\\
&-&\frac{v_F}{\pi a}(|r_1|-|r_2|)\sin\left(\sqrt{2\pi}\phi_c(0)\right)\sin\left(\sqrt{2\pi}\phi_s(0)\right).\nonumber
\end{eqnarray}
We now introduce the charge fluctuation field ${\hat \phi_c(0)}=\phi_c(0)-\sqrt{\pi/2}N$. The total charge in the dot is pinned at its classical value $Q_{cl}=\phi_c(0)=\sqrt{\pi/2}N$ to minimize the Coulomb energy. For weak backscattering $|r_1|,|r_2|\ll 1$ and for energies below the charging energy we can average over the charge fluctuations ${\hat \phi_c(0)}$. When averaging over the term $\cos\left(\sqrt{2\pi}\phi_c(0)\right)=\cos\left(\sqrt{2\pi}{\hat \phi_c(0)}+\pi N\right)$ in Eq.~(\ref{bs3}) we obtain the expression 
\begin{equation}
{\rm e}^{-\pi\langle{\hat \phi_c(0)}^2\rangle}\cos (\pi N)=\sqrt{\frac{a\gamma E_C}{v_F}}\cos(\pi N).
\end{equation}
Here the angular brackets mean averaging with regard to the ground-state of the free Hamiltonian Eq.~(\ref{kinetic2}). An analogous expression can also be found for the term $\sin\left(\sqrt{2\pi}\phi_c(0)\right)$ in Eq.~(\ref{bs3}). When calculating the correlator
\begin{equation}
\langle{\hat \phi_c(0)}^2\rangle=-\frac{1}{2\pi}{\rm ln}\frac{a\gamma E_C}{v_F}
\end{equation}
we have taken into account that only fluctuation modes with energies larger 
than the charging energy $E_C$ can enter into the dot. 
Here $\gamma$ is defined through $\gamma={\rm e}^{\cal C}$ where ${\cal C}\approx 0.577$ is Euler's constant. 
 After the averaging the backscattering part of the Hamiltonian can thus be written as
\begin{eqnarray}\label{bs4}
H_{bs}&=&\frac{\sqrt{\gamma a E_C v_F}}{\pi a}(|r_1|+|r_2|)\cos\left(\pi N\right)\cos\left(\sqrt{2\pi}\phi_s(0)\right)\\
&-&\frac{\sqrt{\gamma a E_C v_F}}{\pi a}(|r_1|-|r_2|)\sin\left(\pi N\right)\sin\left(\sqrt{2\pi}\phi_s(0)\right).\nonumber
\end{eqnarray}
Using simple trigonometric relations the Hamiltonian Eq.~(\ref{bs4}) can be rewritten in an alternative form \cite{m3} as  
\begin{equation}\label{bs5}
H_{bs}=\frac{\sqrt{a\gamma E_C v_F}}{\pi a}\left(r {\rm e}^{i\sqrt{2\pi}\phi_s(0)}+r^\ast{\rm e}^{-i\sqrt{2\pi}\phi_s(0)}\right),
\end{equation}
where $r$ is the complex parameter $r=(|r_1|{\rm e}^{i\pi N}+|r_2|{\rm e}^{-i\pi N})/2$. 
 The first form of the backscattering term Eq.~(\ref{bs4}) will be used in Sec.~\ref{anisotropic} when discussing the two-channel anisotropic version of the Kondo model, while  Eq.~(\ref{bs5}) will be useful as a starting point for the mapping on the isotropic form of the two-channel Kondo model (see Sec.~\ref{isotropic}). The charge part of the kinetic energy can now be dropped since it is completely decoupled from the perturbation due to the backscattering.

\section{Channel-anisotropic two-channel Kondo model}\label{anisotropic}

In this section we will concentrate exclusively on the case  where the reflection amplitudes $|r_1|$ and $|r_2|$ are very close to each other. It is then convenient to  introduce the parameters $|R|=|r_1|+|r_2|$ and $|\delta r|=|r_2|-|r_1|$ where $|\delta r|\ll|R|\ll 1$. For the sake of clarity we have split up this section into three subsections. In the first subsection we will discuss the relation between our problem and the channel-anisotropic (two-channel) 
Kondo model\cite{karyn}.  In the second subsection we will derive an expression for the impurity correction to the ground state energy while a  physical realization of the case $|\delta r|\ll|R|\ll 1$ is discussed in the third subsection. There we  will also explicitly calculate the charge in the dot and the dot's capacitance.

\subsection{Equivalence to the Kondo Model}

 We will now see that our theory for the Coulomb blockade problem can be mapped on the anisotropic two-channel Kondo model at the Emery-Kivelson line
i.e. $J_{\sigma,z}=2\pi v_F$ with $\sigma=1,2$. The solution of 
this model \cite{fabrizio,tsvelik} is due to Fabrizio, Gogolin and Nozi\`eres\cite{fabrizio}.

Introducing the auxiliary impurity-spin operators $\hat{S}_x$ and $\hat{S}_y$ we rewrite the backscattering term  (Eq.~(\ref{bs4})) as 
\begin{equation}\label{bs6}
H_{bs}=\frac{J_x}{\pi a}\cos\left(\sqrt{2\pi}\phi_s(0)\right)\hat{S}_x+\frac{J_y}{\pi a}\sin\left(\sqrt{2\pi}\phi_s(0)\right)\hat{S}_y,
\end{equation}
where the Kondo coupling parameters are defined through 
\begin{eqnarray}\label{coupling}
J_x&=&2|R|\sqrt{a\gamma E_C v_F}\cos(\pi N),\\
J_y&=&2|\delta r|\sqrt{a\gamma E_C v_F}\sin(\pi N).\nonumber
\end{eqnarray}
Starting with a standard Kondo model, we rather get $J_x\propto (J_{1,\perp}+
J_{2,\perp})$ and $J_y\propto (J_{1,\perp}-
J_{2,\perp})$, where $J_{\sigma,\perp}$ denotes the transverse Kondo coupling
of each conduction band channel with the magnetic impurity.
 The total Hamiltonian which we will denote $H^A_{EK}$ is given by $H^A_{EK}=H_{Kin}(\phi_s,\theta_s)+H_{bs}(\phi_s)$, where the kinetic term is of the form Eq.~(\ref{kinetic2}). Both $\hat{S}_x$ and $\hat{S}_y$ can be considered as good quantum numbers since their commutators with the Hamiltonian  $H^A_{EK}$ are small close to perfect transmission through the quantum point contact:
\begin{eqnarray}
\left[H^A_{EK},\hat{S}_x\right]&\propto& -i|R|\hat{S}_z,\\
\left[H^A_{EK},\hat{S}_y\right]&\propto& i|\delta r|\hat{S}_z.\nonumber
\end{eqnarray}
 The impurity spin in $H^A_{EK}$ can oscillate between the two values $\hat{S}_x=1/2$ and $\hat{S}_y=-1/2$. It is important to note that the backscattering part of the total Hamiltonian (Eq.~(\ref{bs4})) and the coupling term of the two-channel anisotropic Kondo model (Eq.~(\ref{bs6})) are exactly equivalent only in the channel symmetric case $|\delta r|=0$, since $\hat{S}_x$ and $\hat{S}_y$ do not commute. However, we will see in the following section that the approximation made in writing Eq.~(\ref{bs6}) is good as long as  $|\delta r|\ll|R|$ or $N\sim 0,\pm 1/2,\pm 1\ldots$.

 We now want to find the shift of the energy  due to the backscattering or, in the language of the Kondo problem, the impurity correction to the ground state energy. To make further progress it is useful to refermionize the Hamiltonian. The basic idea is to introduce a unique operator $\psi(x)$ such that $\cos(\sqrt{2\pi}\phi_s(0))\propto \psi(0)+\psi^\dagger(0)$ and $\sin(\sqrt{2\pi}\phi_s(0))\propto \psi(0)-\psi^\dagger(0)$. Furthermore we use the Majorana representation
\begin{eqnarray}\label{majorana}
\sqrt{2}\hat{S}_x&=&a=(d+d^\dagger)/\sqrt{2},\\
\sqrt{2}\hat{S}_y&=&-b=(d^\dagger-d)/(i\sqrt{2}),\nonumber
\end{eqnarray}
to express the spin-operators through fermionic operators. The $d$ operators ($d=a+ib$) obey $\{d^\dagger,d\}=1$ and  $\{d,\psi(x)\}=0$. The (unusual) 
refermionization procedure will be extensively discussed in Appendix \ref{referm}. There we also give a precise definition of the new fermionic fields $\psi$. After refermionization the kinetic energy takes the standard form
\begin{equation}\label{kinetic5}
H_{Kin}=-iv_F\int^{\infty}_{-\infty}dx\ \psi^\dagger(x)\partial_x\psi(x),
\end{equation}
while the backscattering Hamiltonian is 
\begin{eqnarray}\label{bs7}
H_{bs}&=&\frac{iJ_x}{\sqrt{4\pi a}}\left(\psi(0)+\psi^\dagger(0)\right)b\\&+&\frac{J_y}{\sqrt{4\pi a}}\left(\psi(0)-\psi^\dagger(0)\right)a.\nonumber
\end{eqnarray}
So we finally arrive at a (standard) solvable resonant level type of model for the two Majorana fermions $a$ and $b$~\cite{fabrizio,tsvelik}. 

\subsection{Impurity corrections and scattering phase shifts}

Our next step is to calculate the impurity corrections to the ground state energy from the impurity Green's functions $G_a=-\langle T_\tau a(\tau)a(0)\rangle$ and $G_b=-\langle T_\tau b(\tau)b(0)\rangle$. The Fourier transforms of these  Green's functions can be found in a convenient way from the equations of motion (cf. Appendix \ref{green2}). We get
\begin{equation}\label{green}
G_k(\omega)=\frac{1}{\omega+i\Gamma_k sgn(\omega)},
\end{equation}
where the label $k$ is $k=a,b$. The width of the resonant $a(b)$ level  $\Gamma_a(\Gamma_b)$  is related to the respective Kondo coupling constant $J_y(J_x)$ via
\begin{eqnarray}\label{resonance}
\Gamma_a&=&\frac{{J_y}^2}{4 \pi a v_F}=\frac{E_C\gamma}{\pi}|\delta r|^2\sin^2(\pi N),\\
\Gamma_b&=&\frac{{J_x}^2}{4 \pi a v_F}=\frac{E_C\gamma}{\pi}|R|^2\cos^2(\pi N).\nonumber
\end{eqnarray}
The impurity correction to the ground state energy at zero temperature is
\begin{equation}\label{energy1}
\delta\epsilon=-\int^{E_C}_{\omega_{min}}\frac{d\omega}{2\pi} \omega \left(n_a(\omega)+n_b(\omega)\right).
\end{equation}
The occurrence of a high energy cut-off at $E_C$  in Eq.~(\ref{energy1}) is an intrinsic property of the theory developed so far. The low energy cut-off $\omega_{min}$ needs some additional explanation.  We can distinguish two different situations: In the neighborhood of integer values of $N$ the resonance width $\Gamma_a$ is negligibly small. Already at temperatures of the order of  $\Gamma_b$ spin fluctuations will get pinned, since the coupling constant $J_x$ goes to strong coupling (compare Sec.~\ref{renorm}). 
\begin{figure}
\includegraphics[scale=0.6]{{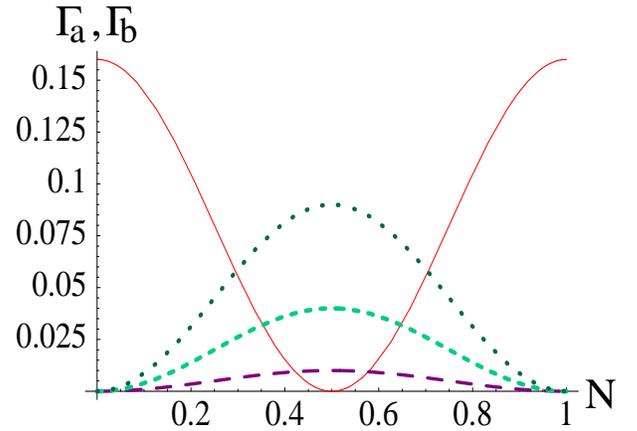}}
\vskip 0.3cm
\caption{\label{resfig}
The resonance energies $\Gamma_a$ and  $\Gamma_b$  of Eq.~(\ref{resonance}) are shown. The parameter $|R|$ in $\Gamma_b$ (full line) is $|R|=0.4$.  The size of the resonance $\Gamma_a$ grows with increasing anisotropy and thus with increasing $|\delta r|$. The lowest (dotted) line in the graph corresponds to  $|\delta r|=0.1$ the second lowest (short dashes) to $|\delta r|=0.2$ and the third (long dashes) to  $|\delta r|=0.3$. The energies are measured in units of $E_C\gamma/\pi$.  }
\end{figure}
 Very close to half-integer values of $N$ however, we have $\Gamma_b\ll\Gamma_a$ even though $|\delta r|\ll|R|$. Thus the coupling $J_y$ will go to strong coupling before $J_x$ and spin fluctuations will be frozen at  $\Gamma_a$. An expression for $\omega_{min}$ that correctly reproduces these two limits is $\omega_{min}=max\{\Gamma_a,\Gamma_b\}$. The density of states $n_k(\omega)$ of impurity $k$ in Eq.~(\ref{energy1})  is related to the corresponding impurity Green's function through
\begin{equation}\label{dos}
n_k(\omega)=-2{\rm Im}\{G_k(\omega)\}=\frac{2\Gamma_k {\rm sgn}(\omega)}{\omega^2+{\Gamma_k}^2}.
\end{equation}
 With this expression for $n_k(\omega)$ the integration in  Eq.~(\ref{energy1}) can easily be performed and we get
\begin{equation}\label{energy2}
\delta\epsilon=-\frac{1}{2\pi}\sum_{k=a,b}\left[\Gamma_k{\rm ln}\left(\frac{{E_C}^2+{\Gamma_k}^2}{\max\{\Gamma_a,\Gamma_b\}^2+{\Gamma_k}^2}\right)\right].
\end{equation}
For most purposes it is sufficient to approximate this expression by the more simple form~\cite{Remark}
\begin{equation}\label{energy3}
\delta\epsilon=-\frac{1}{\pi}\left(\Gamma_a+\Gamma_b\right){\rm ln}\left(\frac{E_C}{\max\{\Gamma_a,\Gamma_b\}}\right).
\end{equation}
We have used $\Gamma_k\ll E_C$ and have dropped an additive constant.  A 
quantity which is interesting is the scattering phase shift (extracted 
from the Friedel
sum rule)
\begin{equation}
\delta (\omega )=\frac{1}{2}\sum_{k=a,b}\arctan \left(\frac{\Gamma_k}{\omega}\right)
\end{equation}
of the conduction electrons $\psi$ due to the impurity scattering. In the channel symmetric case we have $\Gamma_a=0$ and $\delta(\omega \ll \Gamma_b)=\pi/4$. The Friedel sum rule can also be rewritten $Z=2\sum_l\left(2l+1\right)\delta/\pi$ where $Z$ is the impurity charge screened by the electrons. For s-wave scattering $(l=0)$ and $Z=1/2$ we recover $\delta = \pi /4$. The particular value of $\delta$ can thus be understood as a consequence of the fact that only ``half'' of the fermion $d$ (the part $a$) is coupled to the conduction electrons. In general, if both  $\Gamma_a$ and $\Gamma_b$ are finite we find $\delta=\pi/2$ in the limit $\omega\rightarrow 0$ and $Z=1$ since now $a$ and $b$ are screened. 
Here, (even) for a {\it finite} channel asymmetry, we still 
find $\delta=\pi/4$ at the fixed point because close to $N=1/2$ we have
$\Gamma_b\rightarrow 0$ and close to $N=0,\pm 1$ we have $\Gamma_a\rightarrow 0$. This gives a physical justification on the validity of the 
mapping in Sec. IV.
 It is important to remember that the ``spin-fermions'' $\psi$ are not the real electrons but are related to the spin degrees of freedom of the original electronic wave-functions. However, they play the part of the conduction electrons of the real Kondo problem. In the case of spin-less electrons (electrons in a strong magnetic field) and for a single transmitted channel $|r|\ll 1$ the scattering phase shift of the real electrons is related to the average charge on the dot via $\delta=\pi\langle Q\rangle$, again as a consequence of Friedel's rule. It was shown by Aleiner and Glazman\cite{aleiner} that in the one-channel limit this relation can be used to calculate the impurity correction (see Eq.~(\ref{onechannel})) in an intuitive way. 
The physics resembles closely the one of the one-channel Kondo
problem \cite{Nozierephase}.

\subsection{Applications}

 The case $|\delta r|\ll|R|\ll 1$ can be realized in our setup by applying a weak in-plane magnetic field. In a non-zero field the reflection amplitudes $|r_1|$ and $|r_2|$ for electrons with spin-up and spin-down are different due to the Zeeman effect. The channel index $\alpha$ here distinguishes between spin-up and spin-down electrons in the lowest orbital channel $n=1$. The (original) electronic wave-functions for electrons in the two channels are  $\Psi_{\alpha=1}=\Psi^{n=1}_\uparrow$ and $\Psi_{\alpha=2}=\Psi^{n=1}_\downarrow$.  We first want to consider the special case of zero magnetic field which was solved by Matveev in Ref.~\onlinecite{m1}. The reflection amplitudes for the two channels, corresponding to spin-up and spin-down electrons are equal ($|r_1|=|r_2|=|R|/2$) and thus $|\delta r|=0$. It follows from Eqs.~(\ref{coupling}) and (\ref{resonance}) that $J_y=0$ and $\Gamma_a=0$. Furthermore from  Eq.~(\ref{resonance}) we know that $\Gamma_b=|R|^2\gamma E_C\cos^2(\pi N)/\pi$. Calculating the energy with Eq.~(\ref{energy3}) we find\cite{m1}
\begin{equation}
\delta\epsilon=+\frac{\gamma E_C}{\pi^2}|R|^2\cos^2(\pi N){\rm ln}\left(\gamma /\pi |R|^2\cos^2(\pi N)\right).
\end{equation}
The correction to the capacitance can then easily be calculated using Eq.~(\ref{fundamental}) and is found to be 
\begin{equation}
\delta C=-2\gamma E_C|R|^2\beta^2\cos (2\pi N){\rm ln}\left(\frac{1}{|R|^2\cos^2(\pi N)}\right).
\end{equation}
Here we have kept only the logarithmically divergent contribution which will  dominate all other terms close to $N=1/2$. The parameter $\beta=e/(2 E_C)$ is the ratio of the dimensionless parameter $N$ and the gate voltage $V_{G}$, $\beta=N/V_{G}$. 
This seems to be in agreement with the recent capacitance experiment of Ref.~ 
\onlinecite{Ashoori}.
\begin{figure}
\includegraphics[scale=0.6]{{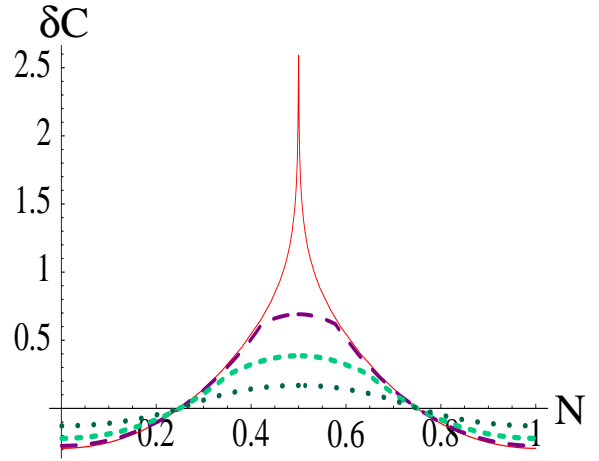}}
\vskip 0.3cm
\caption{\label{capfig}
The correction to the capacitance Eq.~(\ref{anis_cap}) is shown for different values of the anisotropy parameter $|\delta r|=|r_2|-|r_1|$. The parameter  $ |R|=|r_1|+|r_2|$ is set to $|R|=0.4$. The full line corresponds to the channel isotropic case  $|\delta r|=0$ where the capacitance is logarithmically divergent. In order of decreasing  peak height the remaining three curves correspond to $|\delta r|=0.1$, $|\delta r|=0.2$ and  $|\delta r|=0.3$. The units of the capacitance are arbitrary.}
\end{figure}
As we can see from Eq.~(\ref{bs6}) in the channel-symmetric case only fermion $b$ is coupled to the field $\psi$.  If we allow for a weak magnetic field the reflection coefficients for spin-up and spin-down electrons are slightly different and  $|\delta r|\neq 0$. As a consequence the Kondo coupling $J_y$ does not vanish anymore and both Majorana fermions $a$ and $b$ are coupled to the bath.  The total energy shift $\delta \epsilon$ from Eqs.~(\ref{resonance}) and (\ref{energy3}) is found to be\cite{karyn}
\begin{eqnarray}\label{anis_en}
\delta\epsilon&=&+\frac{\gamma E_C}{\pi^2}\left(|R|^2\cos^2(\pi N)+|\delta r|^2\sin^2(\pi N)\right)\\
&\times&{\rm ln}\left(\gamma /\pi\max\{|\delta r|^2\sin^2(\pi N),|R|^2\cos^2(\pi N)\}\right).\nonumber
\end{eqnarray}
Due to the appearance of the max the logarithm does not diverge anymore. Still it can be very large due to the smallness of $|\delta r|$ and the logarithmic term will dominate in the capacitance (See Fig. 3):
\begin{eqnarray}\label{anis_cap}
\delta C&=&-2\gamma E_C\beta^2\left(|R|^2-|\delta r|^2\right)\cos (2\pi N)\\
&\times&{\rm ln}\left(\frac{1}{\max\{|\delta r|^2\sin^2(\pi N),|R|^2\cos^2(\pi N)\}}\right).\nonumber
\end{eqnarray}
We see from  Eqs.~(\ref{anis_en}) and (\ref{anis_cap}) that an arbitrary weak anisotropy between the two reflection coefficients is sufficient to cut off the logarithmic divergence. At the degeneracy point $N=1/2$, we find $\delta C\propto \ln|\delta r|$.
There is some analogy\cite{m1} with the behavior of the 
magnetic susceptibility of the impurity
$\chi=\partial^2(\delta\epsilon)/\partial h^2$
away from the Emery-Kivelson line\cite{fabrizio} and with that of the
local magnetic susceptibility $\chi_l=\partial\langle\hat{S}_z\rangle/\partial h$ 
at the Emery-Kivelson line (the magnetic field
$h$ would only act on the impurity) \cite{tsvelik},
even though for weak backscattering at the QPC there is no real correspondence between the charge in the dot Q and $\hat{S}_z$.
From the two-channel anisotropic Kondo model we see that the appearance of the second energy scale which leads to the suppression of the divergence  follows 
in a natural manner 
from the coupling of the second Majorana fermion to the conduction electrons. 

At this point it is also interesting to compare these results to the result obtained in the case of a single transmitted channel (reflection amplitude $|r_1|\ll 1$
and $|r_2|\rightarrow 1$). Such a situation can be realized in a strong magnetic field \cite{karyn} (spinless electrons) and was also treated in Ref.\onlinecite{m1}. The energy-shift for this case is 
\begin{equation}\label{onechannel}
\delta \epsilon\propto |r_1|E_C\cos (2\pi N)
\end{equation}
and therefore no logarithmic contribution occurs 
anymore in the expressions for the charge and the capacitance: We only
get a periodic oscillation as a function of the gate voltage.  

In the next section we want to demonstrate that the approximation made in writing Eq.~(\ref{bs6}) is justified. Furthermore we will address the case of strong asymmetry between the conduction channels.

\section{Mapping to a channel-isotropic Kondo model}\label{isotropic}

We have seen in the last section that  the backscattering term Eq.~(\ref{bs6}) is not exactly equivalent to the original Hamiltonian Eq.~(\ref{bs4}). We will now justify the approximation made and will rederive the results we have found from the two-channel anisotropic Kondo model in an exact way, following in our derivation Furusaki and Matveev \cite{m3}. The Section is structured in a  similar way as Sec.~(\ref{anisotropic}). The mathematical mapping will be discussed in a first Subsection while a second short Subsection will be devoted to the discussion of an application.

\subsection{Mapping}

 As a starting point we use the form Eq.~(\ref{bs5}) of the original backscattering Hamiltonian which we rewrite as
\begin{equation}\label{bs8}
H_{bs}=\frac{J_0}{2\pi a}\left(r {\rm e}^{i\sqrt{2\pi}\phi_s(0)}+r^\ast{\rm e}^{-i\sqrt{2\pi}\phi_s(0)}\right)\hat{S}_x.
\end{equation}
Here $r$ is the complex parameter $r=(|r_1|{\rm e}^{i\pi N}+|r_2|{\rm e}^{-i\pi N})/2$ and $J_0=4\sqrt{a\gamma E_C v_F}$. Exactly as in the previous section we have introduced an auxiliary impurity spin $\hat{S}_x$. The spin  $\hat{S}_x$ is a good quantum number since it commutes with the total Hamiltonian (here $
\hat{S}_x=1/2$). It is important to note that therefore Eq.~(\ref{bs8}) is exactly equivalent to the original expression Eq.~(\ref{bs5}).  Reformionizing as described in Appendix \ref{referm} gives
\begin{equation}\label{bs9}
H_{bs}=\frac{iJ_0}{\sqrt{4\pi a}}\left(r\psi(0)+r^\ast\psi^\dagger(0)\right)b,
\end{equation}
where we again use the Majorana representation for the impurity spin. The Hamiltonian Eq.~(\ref{bs9}) is very similar to the resonant level model that occurs in the solution of the two-channel isotropic Kondo model at the Emery-Kivelson line (compare Eq.~(\ref{bs7}) with $J_y=0$). The impurity correction to the ground state energy can again be obtained from the Green's function. The propagator for the impurity has exactly the form Eq.~(\ref{green}) with $\Gamma_k$ replaced by 
\begin{equation}\label{resonance2}
\Gamma=\frac{{J_0}^2|r|^2}{4\pi a v_F}=\Gamma_a+\Gamma_b.
\end{equation}
The second equation can be verified by explicit calculation of the sum of the two resonance energies given in Eq.~(\ref{resonance}).
The impurity correction can be found from  Eq.~(\ref{energy1}) in the channel-symmetric limit. Furthermore the density of states of the impurity was defined in Eq.~(\ref{dos}). Combining these expressions we find for the impurity energy
\begin{equation}\label{energy4}
\delta\epsilon=-\frac{\Gamma}{2\pi}{\rm ln}\left(\frac{{E_C}^2+\Gamma^2}{2\Gamma^2}\right)\approx -\frac{\Gamma}{\pi}{\rm ln}\left(\frac{E_C}{\Gamma}\right).
\end{equation}
In the second equation we have used  $\Gamma\ll E_C$. It is now clear that the use of the two-channel anisotropic model is justified whenever ${\rm max} \{\Gamma_a,\Gamma_b\}\approx\Gamma_a+\Gamma_b$. This condition is certainly met when $N$ is very close either to an integer or to a half-integer value. Furthermore the range of $N$-values for which the equation is approximately true will be larger for stronger asymmetry between  $\Gamma_a$ and $\Gamma_b$ i.e. for
a weak asymmetry between the reflection amplitudes $|r_1|$ and $|r_2|$. 

In the two cases of electrons with spin in zero magnetic field or with a weak magnetic field we have ${\rm max}\, (\Gamma_a) \ll{\rm max}\, (\Gamma_b)$ and the channel-anisotropic Kondo model is more convenient (See Fig. 2): In 
particular, the asymmetry simply produces a new energy scale obviously 
affecting the properties of the system close to the degeneracy point $N=1/2$. 
This also allows to make explicit links with the small transmission limit (See
Sec. VII).

\subsection{Applications}

 We will now discuss a case  for  which the isotropic model is particularly well suited. This is the case of strong asymmetry between the reflection amplitudes of the two transmitted channels. We assume that one of the two channels is very close to perfect transmission, e.g. $|r_1|\rightarrow 0$ and $|r_1|\ll |r_2|\ll 1$. Such a situation can be reached in a strong magnetic field where the electrons are essentially spin polarized. The number of open channels and their reflection amplitudes can then be adjusted changing the voltage applied to the gates used to define the QPC. The wave-functions for the electrons in the two channels are  $\Psi_{\alpha=1}=\Psi^{n=1}_{\uparrow (\downarrow)}$ and $\Psi_{\alpha=2}=\Psi^{n=2}_{\uparrow (\downarrow)}$. In the limit of interest here the energy $\delta\epsilon$ with the help of Eqs.~(\ref{resonance2}) and (\ref{energy4}) is found to be
\begin{eqnarray}
\delta\epsilon&=&\frac{\gamma E_C}{\pi^2}|r_2|^2\left(1+2\lambda\cos (2\pi N)\right)\\
&\times&{\rm ln}\left((\gamma |r_2|^2/\pi) \left(1+2\lambda\cos (2\pi N)\right)\right)\nonumber
\end{eqnarray}
where  $\lambda=|r_1|/|r_2|$ is a small parameter. To leading order in  $\lambda$ the correction to the capacitance is given by
\begin{equation}
\delta C=+8\gamma E_C{\beta}^2|r_1||r_2|{\rm ln}\left(|r_2|^2\right)\cos(2\pi N).
\end{equation}
First we observe that at perfect transmission $|r_1|=0$ there is no signature of Coulomb blockade left as it is expected \cite{flensberg,m2,aleiner,nazarov}. Furthermore it is interesting to notice that with one channel very close to perfect transmission the Coulomb blockade oscillations are strongly reminiscent of the one-channel case, also discussed in Ref.\onlinecite{m1}.

\section{Renormalization group formulation}\label{renorm}

We will now show that the energy scales $\Gamma_a$ and $\Gamma_b$ can also be found from a renormalization group treatment. To do this we will use a similar argument as was put forward in Ref.\onlinecite{aleiner} which in turn is based on Ref.\onlinecite{kane}. We will first discuss the channel symmetric case $|r_1|=|r_2|$  (electrons with spin and zero magnetic field). The backscattering Hamiltonian for this case is given by
\begin{equation}\label{bs10}
H_{bs}=\frac{2|r_1|}{\pi a}\sqrt{a\gamma E_C v_F}\cos (\pi N)\cos(\sqrt{2\pi}\phi_s(0)).
\end{equation}
The basic idea of the renormalization group treatment applied here is to calculate the partition function to second order in the small parameter $|r_1|$  and to choose  $|r_1|$ as a function of the lattice step $a$ in such a way that the partition function remains invariant under the transformation $a\rightarrow a'=a\exp(l)$. The correction to the partition function due to the backscattering perturbation is 
\begin{eqnarray}
\delta Z&=&-\int^\beta_0d\tau_1d\tau_2\langle H_{bs}(\tau_1)H_{bs}(\tau_2)\rangle \\
&=&-\frac{{J_x}^2}{(2\pi a)^2}\int^\beta_0d\tau_1d\tau_2\frac{a}{v_F|\tau_1-\tau_2|}.\nonumber
\end{eqnarray}
Using the definition of $J_x$ given in Eq.~(\ref{coupling}) it can be seen that the partition function $\delta Z$ is independent of the lattice step $a$ and thus  $|r_1|$ is invariant under rescaling.  We now introduce a 
{\it dimensionless} parameter $|{\tilde r}_1|$ via $|{\tilde r}_1|=J_x/v_F$. With the help of this parameter we can rewrite the backscattering Hamiltonian 
in the standard form
\begin{equation}\label{bs11}
H_{bs}=\frac{v_F|{\tilde r}_1|}{2\pi a}\cos(\sqrt{2\pi}\phi_s(0)).
\end{equation}
From this we can see that $|{\tilde r}_1|=4|r_1|\sqrt{\gamma E_C a/v_F}\cos(\pi N)$ has the meaning of an effective reflection amplitude. Since $|{\tilde r}_1|\propto \sqrt{a}|r_1|$ it is clear that 
$|{\tilde r}_1|$ grows under the renormalization $a\rightarrow a'=a\exp(l)$. The renormalization flow equation for $|{\tilde r}_1|$ is given by 
\begin{equation}
\frac{d}{dl}{|\tilde r}_1(l)|=\frac{1}{2}|{\tilde r}_1(l)|.
\end{equation}
 We can now integrate this equation from $l=0$ ($a'=a$) to $l=l_c$ ($a'=a_c=a\exp(l_c)$), where $a_c$ is the value of $a'$ for which $J_x\propto|{\tilde r}_1|$ departs to strong coupling, that is  $|{\tilde r}_1(a')|\sim 1$. Integrating we get $|{\tilde r}_1(a_c)|^2=(a_c/a)|{\tilde r}_1(a)|^2=1$. The corresponding critical energy scale is defined through 
\begin{equation}\label{critical}
 E_{x,c}\approx v_F/a_c=v_F|{\tilde r}_1(a)|^2/a={J_x}^2/(a v_F).
\end{equation}
Comparing Eq.~(\ref{critical}) to Eq.~(\ref{resonance}) we see that $E_{x,c}$ is equal to $\Gamma_b$ up to a numerical factor. Further evaluation of  Eq.~
(\ref{critical}) gives
\begin{equation}
E_{x,c}=16|r_1|^2\gamma E_C a\cos^2(\pi N).
\end{equation}
Similar considerations can be used in the more general case where $|r_1|\neq|r_2|$. Then we have to use e.g. the backscattering Hamiltonian Eq.~(\ref{bs4}) as a starting point. Expanding the partition function to the second order in this perturbation (the small parameters are $|R|=|r_1|+|r_2|$ and $|\delta r|=|r_2|-|r_1|$) we get
\begin{equation}
\label{flow}
\delta Z=-\frac{{J_x}^2+{J_y}^2}{(2\pi a)^2}\int^\beta_0d\tau_1d\tau_2\frac{a}{v_F|\tau_1-\tau_2|}
\end{equation}
where the partition function is again independent of the lattice step $a$. Furthermore it is important to note that terms proportional to $J_x$ and $J_y$ do not mix in the expansion of the partition function up to second order.  

For a small channel anisotropy, either the coupling $J_x$ or $J_y$ grows under renormalization which gives rise to two {\it independent} energy scales. The critical energy for which $J_x$ ($J_y$) departs to strong coupling is $E_{x,c}={J_x}^2/(v_Fa)\propto E_C|R|^2\cos^2(\pi N)$ ($E_{y,c}={J_y}^2/(v_Fa)\propto |\delta r|^2 E_C\sin^2(\pi N)$). Of course we have recovered here, up to a numerical factor the two energies $\Gamma_a(\propto E_{y,c})$ and $\Gamma_b(\propto E_{x,c})$ (see Eq.~(\ref{resonance})). 
More generally, using Eq.~(\ref{flow}) we find:
\begin{equation}
\frac{d}{dl}\hbox{\Large{(}}|{\tilde R}(l)|^2+|{\tilde \delta r}(l)|^2
\hbox{\Large{)}}=\hbox{\Large{(}}|{\tilde R}(l)|^2
+|{\tilde \delta r}(l)|^2\hbox{\Large{)}},
\end{equation}
where we defined $|{\tilde R}|=J_x/v_F$ and $|{\tilde \delta r}|=J_y/v_F$. In the case of a strong channel anisotropy, we deduce that the effective
coupling ${J_x}^2+{J_y}^2$ flows off to strong coupling at the critical energy
$E_c=(E_{x,c}+E_{y,c})$. Again, we recover 
$E_c\propto \Gamma=\Gamma_b+\Gamma_a$. For a small anisotropy, of course
this reduces
to $E_c\propto {\rm max} \{\Gamma_a,\Gamma_b\}$.

\section{Realization in a two-contact set-up}

Let us now briefly 
emphasize that the channel-anisotropic two-channel Kondo model 
of Sec. III is also well suited to discuss  the
conductance behavior of a 
two-contact set-up \`a la Furusaki-Matveev. In this  setup which is a natural
extension of the geometry shown in Fig.\ref{dot} a single dot is
coupled to two different electron reservoirs via two point contacts. A
back-gate or side-gate is used to vary the Coulomb energy.  
The device is illustrated in Ref.\onlinecite{m3}. 

Let us
start with a strong magnetic field such that the electrons
 of each reservoir are
fully  polarized. We obtain an effective two-channel model where 
$\alpha=1(2)$ now denotes the electrons of the left (right) QPC. Again, we 
are mainly interested to the case of a
{\it small channel anisotropy} where reflection amplitudes at 
each contact, namely $|r_1|$ and $|r_2|$, are close to each 
other: $|r_1|<|r_2|\ll 1$.
It is convenient to associate the centers of the two constrictions
with the point
$x=0$, meaning that electrons in the dot are described by $\Psi_{1,L(R)}$
at $x>0$ and by $\Psi_{2,L(R)}$ at $x<0$\cite{m3}.
We introduce the variables $\phi_t$ and $\phi_c$ via
\begin{equation}
\phi_1=\frac{1}{\sqrt{2}}(\phi_t+\phi_c),\qquad 
\phi_2=\frac{1}{\sqrt{2}}(\phi_t-\phi_c),
\end{equation}
and similarly for $\theta_1$ and $\theta_2$. 
Typically, here $\phi_c$ is the bosonic variable associated to the {\it charge}
in the dot, while $\phi_t$ is the one related to the {\it current} passing 
through the dot. The main difference with the
one-contact set-up is that here the charge on the dot rather reads
(note the minus sign)\cite{m3}
\begin{equation}
Q=\frac{1}{\sqrt{\pi}}\hbox{\large{(}}\phi_1(0)-\phi_2(0)\hbox{\large{)}}.
\end{equation}
It is straightforward to
show that the backscattering Hamiltonian is of the form of 
Eq. (16) or Eq. (18) with $\phi_s$ replaced by $\phi_t$, explicitely
\begin{equation}\label{an-two}
H_{bs}=\frac{J_x}{\pi a}\cos\left(\sqrt{2\pi}\phi_t(0)\right)\hat{S}_x+\frac{J_y}{\pi a}\sin\left(\sqrt{2\pi}\phi_t(0)\right)\hat{S}_y.
\end{equation}
The Kondo parameters are  
given in Eq.~(\ref{coupling}) where $|\delta r|\ll |R|\ll 1$,
and the kinetic energy for the symmetric charge mode
 $H_{Kin}(\phi_t,\theta_t)$ is still of the form of Eq.~(\ref{kinetic2}).
Again, we insist on the fact that this mapping is more intuitive than the one
of Eq.~(\ref{bs5}) 
because the anisotropy of the reflection coefficients is directly
related to the anisotropy between coupling parameters in the Kondo model. In
consequence, we have two independent energy scales $\Gamma_a$ and $\Gamma_b$.
The behavior of the conductance close to the degeneracy point $N=1/2$ 
becomes really transparent. 

In the absence of any backscattering, the
system is equivalent to two resistances $2\pi\hbar/e^2$ (those of the two QPCs)
connected in series. Therefore the conductance is $G_0=e^2/(4\pi\hbar)$. 
This is in fact still the case for {\it finite} barriers in the symmetric case
$|r_1|=|r_2|$ at the degeneracy point\cite{Nagaosa}, where both $J_{x}$ and
$J_y$ are zero for $N=1/2$. As  was found in Ref.\onlinecite{m3}
the conductance is still at its resonant value $G_0$ 
and the tails of the peak at $N=1/2$ are not
Lorentzian: $G\propto (N-1/2)^{-4}$ due to the $J_x$ coupling. 
When $|\delta r|$ is finite, from Eq. (52) we immediately see that
the asymmetry between channels engenders a finite backscattering process at
$N=1/2$ (again, $J_{y}\propto |\delta r|$):
\begin{eqnarray}
H_{bs} &=& \frac{J_y}{\pi a}\sin\left(\sqrt{2\pi}\phi_t(0)\right)\hat{S}_y,
\\ \nonumber
&=& \frac{J_y}{\sqrt{4\pi a}}\left(\psi(0)-\psi^\dagger(0)\right)a.
\end{eqnarray} 
This produces the same physics as a non-magnetic
impurity in a Luttinger liquid with the Luttinger exponent being $g=1/2$
\cite{kane}.
The problem becomes exactly solvable using the refermionization 
procedure of Appendix A.  
In particular, the current through the device
takes the required form\cite{tsvelik}:
\begin{equation}
I=ev_F\psi^{\dagger}(0)\psi(0).
\end{equation}
For temperatures $T\ll E_{y,c}$ and $N$ close to 1/2, the effective 
potential scattering $J_y$ (or the asymmetry parameter) 
{\it diverges} and the conductance then obeys\cite{tsvelik}:
\begin{equation}
G(T;N\approx 1/2)=G_0\hbox{\huge{(}} \frac{T}{E_{y,c}}\hbox{\huge{)}}
^{2/g-2}\propto
\hbox{\huge{(}} \frac{T}{E_c N^2}\hbox{\huge{)}}^{2},
\end{equation}
where the second equation is true for $g=1/2$. The energy $E_{y,c}$
was defined in the last paragraph of the previous section.
 For non-symmetric barriers the conductance peak height (like the capacitance
peak height before) becomes strongly dependent on the finite 
asymmetry between reflection amplitudes at the QPCs (Fig. 4). This
naturally 
reflects the fact that the on-resonance behavior of the system for $N=1/2$ 
is very different if the two 
barriers are not identical. 

A simple explanation can be given using
 the channel-anisotropic two-channel Kondo model. On resonance, the
small asymmetry inevitably grows under renormalization. 
At low temperatures $T\ll {\rm max} \{\Gamma_a,\Gamma_b\}$, the conductance behavior for all values of $N$ is given by
\begin{equation}
G(T;N)=G_0\hbox{\huge{(}}
\frac{T}{{\rm max} \{\Gamma_a,\Gamma_b\}}\hbox{\huge{)}}^{2}\ll G_0.
\end{equation}
This can also be rewritten as\cite{m3}
\begin{equation}
G(T;N)\propto\hbox{\huge{(}}
\frac{T}{\Gamma}\hbox{\huge{)}}^{2},
\end{equation}
where (See Eq. (38))
\begin{equation}
\Gamma\propto \hbox{\Large{(}}  |r_1|^2+|r_2|^2+2|r_1||r_2|\cos(2\pi N)
\hbox{\Large{)}},
\end{equation}
even though this formula seems a little bit less intuitive. In particular, the
N-dependence of the conductance for $N\approx 1/2$ is less apparent.
It is worth noting that the quadratic temperature dependence is a universal 
property of inelastic co-tunneling. But, close to $N=1/2$
this can also be interpreted as 
a nice manifestation of the restoration of the {\it Fermi-liquid} behavior 
due to the finite
\vskip -0.3cm
\begin{figure}[ht]
\centerline{\epsfig{file=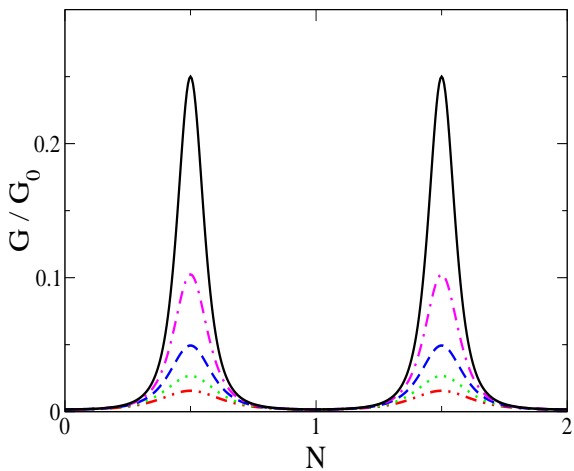,angle=0.0,height=7.7cm,width=8.7cm}}
\vskip -0.2cm
\caption{Conductance in the two-contact set-up (in strong magnetic field) 
as a function of $N$ 
for different values of the {\it (small)}
anisotropy parameter $|\delta r|=|r_2|-|r_1|$. 
The parameter $|R|=|r_1|+|r_2|$ is here set to $|R|=7/10$. The different curves
 have been calculated for $T/E_c=1/50$ and have been normalized to
$G_0$. In order of decreasing conductance peak height the different curves here
correspond to $|\delta r|=0.2$ (top), $|\delta r|=0.25$, $|\delta r|=0.3$, $|\delta r|=0.35$ and $|\delta r|=0.4$ (bottom).}
\end{figure}
asymmetry between channels (like for the capacitance problem).

Unfortunately, the present approach does not allow a complete solution for 
the case of fermions with spin in the two-contact set-up\cite{m3}. At zero 
magnetic field
and for completely symmetric barriers at the two QPCs, there exists  
a mapping onto a four-channel Kondo model for the Hamiltonian (which
could help to compute thermodynamical properties), but it is 
very difficult to rigorously compute current-current
correlation functions in the new basis and then to extract the conductance
behavior; The current operator indeed exhibits
an unusual form as indicated in Appendix B of Ref.~\onlinecite{m3}. 
Therefore, the height of the conductance peak
cannot be calculated in a rigorous manner even for completely
symmetric barriers at zero temperature. 
Furthermore,  the mapping onto the Kondo
problem is only valid at (close to) $N=1/2$. 
To extend the (four-channel) Kondo mapping to the case of (weakly) asymmetric
barriers remains a challenging task.      

\section{Small Transmission Limit}\label{sec_t_small}

Having so far concentrated on the limit where the quantum dot is strongly coupled to a reservoir through a highly transmissive quantum point contact, we will in this section consider the limit of weak coupling. The similarity between the Coulomb blockade problem in this limit and a Kondo model was noticed  by Glazman and Matveev \cite{glazman}. An explicit mapping of the Coulomb blockade Hamiltonian on a Kondo model was used by Matveev to calculate the charge and the capacitance of the quantum dot in the weak transmission limit\cite{m2}.  
Note that recently a noncrossing approximation (NCA) has been generalized to 
this type
of multichannel Kondo models \cite{Schiller}.

We will here rederive Matveev's mapping and discuss the straight-forward extension of his model to the channel anisotropic case. Again our discussion will be restricted to the case of two transport channels through the point contact.
The anisotropy between the transmission amplitudes for the two transport channels in the QPC will still give rise to a mapping on a channel-anisotropic Kondo model; But here, the system flows off to the usual {\it spin-isotropic} fixed
point, i.e., $J_{\sigma,\perp}=J_{\sigma,z}\gg 0$. We will see below that  
the problem of calculating the average charge $\langle Q\rangle$ on the dot is now equivalent to the problem of finding the average of the $z$-component $\langle S_z\rangle$ of the impurity spin in the Kondo model. In addition the capacitance of the dot is the same as the  magnetic susceptibility $\chi$ of the impurity. Once this equivalence is established the problem is solved, since these quantities ($\langle S_z\rangle$, $\chi$) can be found in the literature. The channel- and spin {\it isotropic} multichannel Kondo problem was solved exactly in this limit in Refs.~\onlinecite{wiegmann,andrei}, while the 
impurity-susceptibility for the {\it channel-anisotropic} (but spin-isotropic) 
two-channel Kondo problem can be simply extracted from
Ref.~\onlinecite{coleman}. 
Note in passing that the channel-anisotropic case was also  
solved (exactly) using Bethe Ansatz in Ref.~\onlinecite{andrei2}; There, 
the Wilson
ratio is computed even in the case of a channel anisotropy. This model has
also been investigated using conformal field theory and numerical 
renormalization-group calculations in Ref.~\onlinecite{affleck-ani}.

We will now proceed to writing down the model for our system in the small transmission limit. Instead of formulating it in momentum space as it was done by Matveev we will here use a real space formulation which allows us to stress the analogy with the corresponding model in the strong transmission limit (see Sec.~\ref{model}). In Appendix \ref{t_small}, we will also address 
the question why the mapping on the Kondo model can not be derived using bosonization as it was done in Sec.~\ref{anisotropic}. In the previous section we have treated the backscattering as a small perturbation to an otherwise perfectly transparent QPC. 
In this section the perturbation is a tunneling Hamiltonian which couples two a priori independent systems (the 2DEG and the dot). A smooth transition can be made from the strongly to the weakly coupled limit by continuously increasing the auxiliary gate voltage to pinch off the QPC. In this transition a perfectly transmissive one-dimensional channel $n$ will be cut into two weakly coupled halves. In the vicinity of the center of the QPC electron motion is still quasi one-dimensional. Electronic wave-functions to the left of the center of the QPC (in the reservoir) will be denoted $\Psi^n_{\sigma,0}$, wave-functions to the right of the center (in the QD) are $\Psi^n_{\sigma,1}$. Again $\sigma$ is the spin index, $n$ the channel index due to lateral confinement and here $\alpha=0,1$ indicates the location of the electron. Hopping between the two sides of the QPC constitutes a small perturbation. To model this perturbation we use the Hamiltonian
\begin{equation}\label{tunneling}
H_{T}=\sum_{\sigma=1,2}\left(|t_\sigma| \Psi^\dagger_{\sigma,1}(0)\Psi_{\sigma,0}(0)+h.c.\right).
\end{equation}
Note that the kinetic energies  $H^0_{kin}\left(\{\Psi^\dagger_{\sigma,0},\Psi_{\sigma,0}\}\right)$ and $H^1_{kin}\left(\{\Psi^\dagger_{\sigma,1},\Psi_{\sigma,1}\}\right)$ for electrons in the 2DEG and the dot have the form of Eq.~(\ref{kinetic5}). The boundaries for the integration along the $x$-axis are $-\infty, 0$ in $H^0_{kin}$ and $0,\infty$ in $H^1_{kin}$. The Coulomb interaction can be modeled as in Sec.~(\ref{model}), Eq.~(\ref{coulomb1}) the charge on the dot being
\begin{equation}
Q=\int^\infty_0 dx \left (\Psi^\dagger_{\sigma,1}(x)\Psi_{\sigma,1}(x)-\rho_0\right).
\end{equation}
The equilibrium charge density $\rho_0$ is chosen in such a way that the total charge $Q$ on the dot is zero when no voltage is applied to the gate  ($V_G=0$). 

We now want to concentrate on the point $N=1/2$ where the states with $Q=0$ and $Q=1$ are energy degenerate (See Eq.~(\ref{coulomb1})), and therefore
charge fluctuations are large. We introduce the small parameter $U=e/(2C_{gd})-V_G\propto (N-1/2)$ to measure deviations of N from the degeneracy point. In terms of $U$ the electrostatic energies of the two states  $Q=0$ and $Q=1$ are $E_0=0$ and $E_1=eU$ respectively. 

If the two conditions $|U|\ll e/C_{gd}$ and $kT\ll E_C$ are met  only the two states  with  $Q=0$ and $Q=1$ are accessible and higher energy states can be removed from our theory introducing the projection operators $P_0$ and $P_1$. Here $P_0$ and $P_1$ are  projecting on the states $Q=0$ and $Q=1$ respectively. The effective Hamiltonian for this truncated system is
\begin{eqnarray}\label{effective}
H_{eff}&=&\left(H^0_{kin}+H^1_{kin}\right)\left(P_0+P_1\right)+eUP_1\\
&+&\sum_{\sigma=1,2}\left(|t_\sigma| \Psi^\dagger_{\sigma,1}\Psi_{\sigma,0}P_0+|t_\sigma| \Psi^\dagger_{\sigma,0}\Psi_{\sigma,1}P_1\right),\nonumber
\end{eqnarray}
where the operators in the tunneling part are evaluated at $x=0$ (see Eq.~\ref{tunneling}). From this equation we see that $\partial H/\partial U=eP_1$ and therefore $\langle Q\rangle = \partial E_0/\partial U$ where $E_0$ is the ground-state energy of the Hamiltonian Eq.~(\ref{effective}). It can be shown that the Hamiltonian Eq.~(\ref{effective}) is equivalent to the Hamiltonian
\begin{eqnarray}\label{kondo}
H_{K}&=&\sum_{\alpha=0,1}H^\alpha_{kin}+2hS_z\\
&+&\frac{1}{2}\sum_{\sigma=1,2}\left(J_{\sigma,\perp}\left(s^+_{\sigma}(0)S^- +s^-_{\sigma}(0)S^+ \right)\right)\nonumber
\end{eqnarray}
which has the form of a standard Kondo Hamiltonian. 
The equivalence between the Hamiltonians Eq.~(\ref{effective}) and  Eq.~(\ref{kondo}) was demonstrated by Matveev in Ref.~\onlinecite{m1}. For completeness we rederive this relation in Appendix \ref{t_small} taking channel anisotropy into account explicitely.
In Eq.~(\ref{kondo}) we have introduced the transversal Kondo coupling parameters $J_{\sigma,\perp}=2|t_\sigma|$ and  an  effective external magnetic field $h=eU/2\propto (N-1/2)$  applied along the $z$ axis.  Note that Eq.~(\ref{kondo}) corresponds to the limit $J_{1,z}=J_{2,z}=0$ of Eq.~(\ref{kondo0}). We have dropped the constant $eU/2$ from the Hamiltonian in Eq.~(\ref{kondo}). Therefore the ground state  energy $E_{K,0}$ of the Kondo Hamiltonian  Eq.~(\ref{kondo}) is related to the ground state  energy $E_{0}$ of the effective Hamiltonian Eq.~(\ref{effective}) through  $E_{K,0}+eU/2=E_0$. Since $2\langle S_z\rangle=\partial E_{K,0}/\partial h$ and  $\langle Q\rangle=\partial E_0/\partial U$ we are led to the obvious identification 
\begin{equation}
\langle Q\rangle=e\left(\frac{1}{2}+\langle S_z\rangle\right).
\end{equation}
Combining Eq.~(\ref{fundamental}) with $U=e/(2C_{gd})-V_G$ we find for the correction to the capacitance $C=-\partial \langle Q\rangle /\partial U \propto \chi=\partial \langle S_z\rangle/\partial h$, where $\chi$ is the impurity susceptibility (Since $h$ only acts on $S_z$, $\chi$ is also equivalent to the
local magnetic susceptibility $\chi_l$). 

So we have now shown that a calculation of the capacitance of the quantum dot coupled to a 2DEG through a QPC with two transport channels with {\it different} transmission amplitudes is equivalent to a calculation of the impurity susceptibility in the channel-anisotropic two-channel Kondo model. It is known that both $J_{\sigma, z}=0$ and $J_{\sigma, \perp}\ll 1$ grow under renormalization and that for small enough energies $J_{\sigma, z}=J_{\sigma, \perp}=J_{\sigma}\propto |t_{\sigma}|$ even if we start from a spin-anisotropic Kondo model. The resulting spin-isotropic Kondo model was solved in Refs.~\onlinecite{wiegmann,andrei}. In the zero temperature limit the impurity susceptibilities for the one-channel case and the two-channel isotropic limit are
\begin{equation}\label{chi}
\chi \propto \left[
       \begin{array}{ll}
       cst. & J_2=0,\\
        \log h & J_1=J_2.
       \end{array}
       \right.
\end{equation}
Note that the susceptibility for the two-channel case diverges in the zero magnetic field limit. The channel-anisotropic (but spin-isotropic) 
Kondo model was discussed e.g. by Coleman and Schofield\cite{coleman} 
(also by Andrei and Jerez\cite{andrei2} and by Affleck 
{\it et al.}\cite{affleck-ani})
who found for the susceptibility at zero magnetic field ($h=0$)
\begin{equation}\label{chianis}
\chi \propto 1-\frac{\log (\nu)}{\nu-1},
\end{equation}
where the anisotropy parameter  $\nu \propto |\delta t|/{\cal T}$ 
with $|\delta t|\propto |J_1-J_2|$ and ${\cal T}\propto (J_1+J_2)$,
is $\nu = 1$ in the one-channel limit ($J_2=0$) and  $\nu = 0$ in the two-channel isotropic limit. Eq.~(\ref{chianis}) thus reproduces correctly the main characteristics of the result Eq.~(\ref{chi}) in these two limiting situations. From the results for the susceptibility we immediately obtain the capacitance
\begin{equation}\label{capacitance}
\delta C \propto \left[
       \begin{array}{ll}
       cst. & J_2=0,\\
       \log ({N-1/2}) & J_1=J_2,\\
        \log|\delta t| & J_1\neq J_2, N=1/2.
       \end{array}
       \right.
\end{equation}
The results for the one-channel and the two channel isotropic cases are due to Matveev\cite{m1}. The main observation we want to make here is that a small channel anisotropy cuts off the logarithmic divergence exactly in the same way as in the case discussed in the previous sections where the reflection amplitudes in the QPC could be treated as small parameters.

\section{conclusions}

We have applied the channel- (and spin-) anisotropic two-channel Kondo model to study Coulomb blockade oscillations in the capacitance of a quantum dot.  Our main interest has been to investigate the effect of an asymmetry between the reflection (or transmission) amplitudes of different open channels in the QPC connecting the dot to a reservoir. Following Matveev\cite{m2,m1} we have studied the two exactly soluble limits of very weak  ($|t_1|,|t_2|\ll 1$) and very strong  coupling ($|r_1|, |r_2|\ll 1$). 

A summary of the results for the capacitance in different limits is given in Table 1. 

In both limits a mapping of the original problem onto a Kondo model is possible.  Remember that these results concern the limit of very low-temperature where quantum fluctuations are prominent. At quite high 
temperature, the Kondo physics gets destroyed by
thermal fluctuations \cite{m1,aleiner,aleiner2}. 

For weak backscattering at the point contact ($|r_1|, |r_2|\ll 1$) the original problem can be mapped\cite{karyn} on a channel-anisotropic Kondo model at the Emery-Kivelson\cite{emery} line ($J_{1,z}=J_{2,z}=2\pi v_F$). For this particular value of the  coupling constant the Kondo model is exactly solvable. The anisotropy of the reflection amplitudes for different channels is directly reflected in the channel-anisotropy of the Kondo model. In fact we found $J_{1,\perp}\propto |r_1|$ and $J_{2,\perp}\propto |r_2|$, where $J_{1,\perp}$ and $J_{2,\perp}$ are the coupling constants for the two channels in the Kondo model. The mapping allowed us to calculate the shift of the ground-state energy of our Hamiltonian due to the backscattering at the point contact and in turn to find the capacitance of the quantum dot. While the capacitance is logarithmically divergent for values of the gate voltage close to $N=n+1/2$ ($n$ is an integer) in the isotropic limit\cite{m1} ($|r_1|=|r_2|$), this divergence is {\bf cut off} by  a small anisotropy between the reflection amplitudes of different channels\cite{karyn}. 

{\it This can be interpreted as a manifestation of the
restoration of the Fermi-liquid
behavior close to the degeneracy points $N=n+1/2$ due to the asymmetry between
channels.}

Note that a similar 
conclusion can be reached investigating the conductance
behavior in a two-contact set-up in a strong magnetic field where 
$|r_1|$ and $|r_2|$ denote the backscattering amplitudes at the two
different point contacts. 
The on-resonance behavior $(G=G_0=e^2/2h)$ for $N=1/2$ is {\it reduced} for a small 
anisotropy between $|r_1|$ and $|r_2|$.

There are two intrinsic energy  scales ($\Gamma_b\propto (|r_1|+|r_2|)^2$ and $\Gamma_a\propto (|r_1|-|r_2|)^2$) in the channel-anisotropic Kondo model. Each can be interpreted as the resonance energy related to coupling half an impurity to the conduction electrons. In the channel-isotropic case only half the impurity spin is screened by the conduction electrons. Coupling back the second half of the impurity to the conduction electrons leads to the emergence of a second energy scale $\Gamma_a$. It is this new energy scale that enters the expression for the capacitance and cuts off the divergence at $N=n+1/2$. Unfortunately there is no direct correspondence between the magnetic susceptibility in the Kondo model and the capacitance of the quantum dot. Such an equivalence exists only in the limit of small transmission. It is true however that the behavior of the capacitance (found from the Kondo model at the Emery-Kivelson line) is reminiscent of the impurity susceptibility $\chi=\partial^2 (\delta \epsilon)/\partial h^2$ away from the Emery-Kivelson line or of the local magnetic susceptibility $\chi_l=\partial \langle \hat{S}_z\rangle/\partial h$ at the Emery-Kivelson line. 

In this paper we have extended our previous work\cite{karyn} deriving the mapping on the anisotropic Kondo model in a  pedagogical way and carefully discussing its limits of validity. We have then given an alternative way for calculating the charge on the quantum dot  using a mapping\cite{m3} on a channel-isotropic two-channel Kondo model. In this approach the coupling constant is a complex parameter depending on both  $|r_1|$ and $ |r_2|$. While the latter approach has the advantage of being exact, it seems less intuitive since the anisotropy of the reflection coefficients is not directly reflected as an anisotropy between coupling parameters in the Kondo model. We have in addition used a simple scaling argument to recover the intrinsic energy scales $\Gamma_a$  and $\Gamma_b$ that occur in the expression of the capacitance.

We relied on purely mathematical arguments to show the equivalence between the  Coulomb blockade problem and the Kondo Hamiltonian in the strong tunneling limit ($|r_1|, |r_2|\ll 1$). In the opposite limit $|t_1|,|t_2|\ll 1$  the similarity of these two problems can be understood using a comparably simple physical argument (see Ref.\onlinecite{m1} and Appendix \ref{t_small}). The main observation is that at low enough temperatures $T\ll E_C$ and for voltages close to  $N=n+1/2$ only two charge states on the quantum dot are energetically accessible (e.g. $Q=0$,  $Q=1$). The charge state of the dot  is then interpreted as a pseudo-spin-$1/2$ degree of freedom which corresponds to the impurity spin in the Kondo model. The real spin of the conduction electrons in the Kondo model is replaced by an index $\alpha$ indicating the location of an electron (distinguishing between electrons to the left and to the right of the QPC) in the Coulomb blockade problem.  

A first order process in the Kondo problem that flips the impurity spin from  up to down and the spin of a conduction electron from down to up is then equivalent to a tunneling process that takes an electron from the left to the right and changes the charge on the dot from  $Q=0$ to  $Q=1$. The Kondo coupling parameters simply are $J_{1(2),\perp}\propto |t_{1(2)}|$ and $J_{1(2),z}=0$.  As it is clear from the derivation of the mapping  there exists an equivalence between the charge on the dot and the impurity spin, namely $\langle Q\rangle \propto \langle S_z\rangle$. It was furthermore shown that the capacitance is basically the same (up to some constant) as the impurity susceptibility. Since the model with $J_{1(2),\perp}\propto |t_{1(2)}|$ and $J_{1(2),z}=0$ flows to the usual spin-isotropic fixed point we have been able to use the known result for the impurity susceptibility in the channel-anisotropic two-channel Kondo model with $J_{\sigma,\perp}=J_{\sigma,z}\gg 0$\cite{coleman}, to discuss the effect of channel-anisotropy on the capacitance. Exactly as in the limit of strong transmission  the capacitance diverges in the channel-isotropic case\cite{m2}, but the divergence is cut by the anisotropy. 
Note that no immediate connection can be made with the Kondo model at the Emery-Kivelson line ($J_{1(2),z}=2\pi v_F$) which we obtained when treating the small reflection limit.  

Let us finally remind that all the results on the behavior of the
capacitance of the dot close to $N=1/2$ have been summarized in 
Table 1.

\acknowledgments{This work was supported by the Swiss National Science Foundation.}

\appendix

\section{refermionization}\label{referm}

In this Appendix we want to elaborate on the unusual refermionization procedure which we used in Sec.~(\ref{anisotropic}). The backscattering  part of the bosonic Hamiltonian to be refermionized is given in Eq.~(\ref{bs6}). The kinetic energy is of the form
\begin{equation}\label{kineticA1}
H_{Kin}=v_F\int^{\infty}_{-\infty}dx\left[\left(\partial_x\phi_s (x)\right)^2+\pi_s(x)^2\right].
\end{equation}
It will turn out to be convenient to use the field  $\pi_s(x)$ instead of $\theta_s(x)$ for the moment. The two fields are related through $\pi_s(x)=\partial_x\theta_s(x)$. The commutation relations for the fields $\phi_s(x)$ and   $\pi_s(x)$ are $[\phi_s(x),\pi_s(y)]=i\delta(x-y)$, $[\phi_s(x),\phi_s(y)]=0$ and $[\pi_s(x),\pi_s(y)]=0$. The basic idea of the refermionization procedure is to introduce an operator $\psi(x)$ such that $\cos(\sqrt{2\pi}\phi_s(0))\propto \psi(0)+\psi^\dagger(0)$ and $\sin(\sqrt{2\pi}\phi_s(0))\propto \psi(0)-\psi^\dagger(0)$. It is clear that for such an operator
\begin{equation}
 \psi(0)=\frac{1}{\sqrt{2\pi a}}\exp (i\sqrt{2\pi}\phi_s(0)).
\end{equation}
 In addition the operator must obey the usual fermionic anti-commutation relations. Using the relation ${\rm e}^A{\rm e}^B={\rm e}^B{\rm e}^A{\rm e}^{[A,B]}$ it can be seen that the obvious choice for $\psi$, namely $\psi(x)=(2\pi a)^{-1/2}\exp (i\sqrt{2\pi}\phi_s(x))$ does not obey anti-commutation relations. To construct a fermionic operator we introduce the auxiliary fields 
\begin{eqnarray}
\phi_\pm(x)&=&\frac{1}{\sqrt{2}}\left(\phi_s(x)\pm\phi_s(-x)\right),\\
\pi_\pm(x)&=&\frac{1}{\sqrt{2}}\left(\pi_s(x)\pm\pi_s(-x)\right).\nonumber
\end{eqnarray}
In terms of these new fields the kinetic energy is
\begin{equation}\label{kineticA2}
H_{Kin}=v_F\sum_{\alpha=\pm}\int^{\infty}_{0}dx\left[\left(\partial_x\phi_\alpha \right)^2+\left(\partial_x\theta_\alpha\right)^2\right].
\end{equation}
We thus arrive at a theory which is confined to positive values of $x$. The advantage of this restriction becomes clear when we  introduce the two additional right-going and left-going fields
\begin{eqnarray}\label{bosefields}
\Phi_{R,\pm}(x)&=&\phi_\pm(x)-\int^x_0dy\ \pi_\pm(y),\\
\Phi_{L,\pm}(x)&=&\phi_\pm(x)+\int^x_0dy\ \pi_\pm(y)).\nonumber
\end{eqnarray}
At $x=0$ we  have $\Phi_{R,+}(0)=\Phi_{L,+}(0)=\sqrt{2}\phi_s(0)$ and also $\Phi_{R,-}(0)=\Phi_{L,-}(0)=0$ which makes these  fields candidates for the construction of our new fermions. The backscattering Hamiltonian can  be expressed through the fields $\Phi_{R,+}(x)$ and $ \Phi_{L,+}(x)$ only. There are in fact many ways in which this can be done. However, we will  soon get rid of this ambiguity. The fields  $\Phi_{R(L),-}(x)$ occur only in the kinetic part of the Hamiltonian and are thus of little interest to us. Later on we will  need the commutation relations 
\begin{eqnarray} 
 \left[\Phi_{R,+}(x),\Phi_{R,+}(y)\right]&=&+i{\rm sgn}(x-y),\\
 \left[\Phi_{L,+}(x),\Phi_{L,+}(y)\right]&=&-i{\rm sgn}(x-y).\nonumber
\end{eqnarray}
 For completeness we also give the correlation functions $G_{R(L)}=\langle\Phi_{R(L),+}(x)\Phi_{R(L),+}(0)-\Phi_{R(L),+}(0)^2\rangle$ which take the standard form
\begin{equation}
G_{R(L)}=\frac{1}{\pi}{\rm ln}\left(\frac{a}{a\pm ix}\right).
\end{equation}
The plus sign belongs to the label $R$ while the minus sign belongs to $L$. 
 The kinetic energy in terms of these new fields takes the  form 
\begin{equation}\label{kineticA3}
H_{Kin}=\frac{v_F}{2}\sum_{\alpha=\pm}\int^{\infty}_{0}dx\left[\left(\partial_x\Phi_{R,\alpha}\right)^2+\left(\partial_x\Phi_{L,\alpha}\right)^2\right].
\end{equation}
 We now drop the  $\Phi_{R(L),-}(x)$ part of the kinetic energy since it is not coupled to the backscattering term. To refermionize the  $\Phi_{R(L),+}(x)$ part we introduce the  operators
\begin{eqnarray}\label{wf1}
\psi_{R}(x)&=&\frac{1}{\sqrt{2\pi a}}{\rm e}^{i\sqrt{\pi}\Phi_{R,+}(x)},\\
\psi_{L}(x)&=&\frac{1}{\sqrt{2\pi a}}{\rm e}^{i\sqrt{\pi}\Phi_{L,+}(x)}.
\end{eqnarray}
Using the commutation  relations for the bosonic fields we can verify that the fields $\psi_{R,\pm}(x)$ and $\psi_{L,\pm}(x)$ really are fermions and obey
\begin{equation}
 \psi_{p,+}(x)\psi_{p,+}(y)=-\psi_{p,+}(y)\psi_{p,\pm}(x),
\end{equation}
where $p=R,L$. Note, that for $x=0$  we have $\psi_{L,+}(0)=\psi_{R,+}(0))$ (see Eqs.~(\ref{bosefields}) and (\ref{wf1})) and there thus seems to be more than one way to refermionize the backscattering Hamiltonian Eq.~(\ref{bs6}). To lift this ambiguity we reextend our theory on the full $x$-axis via the definition
\begin{equation}\label{wf2}
\psi(x)={\cal P}\left[ 
\begin{array}{ll}
\psi_{R}(x) & x>0,\\
\psi_{L}(-x) & x<0.
\end{array}
\right.
\end{equation} 
 In the above definition of the fermion $\psi(x)$ (Eq.~(\ref{wf2})) we have introduced an additional phase factor
\begin{equation}\label{phasefactor}
{\cal P}=\exp (i\pi d^\dagger d)=1-2d^\dagger d
\end{equation}
to ensure that  $\psi(x)$ anticommutes with the spin operators $\hat{S}_x$ and $\hat{S}_y$ written in terms of Majorana fermions $d$ and $d^\dagger$ (cf. Eq.~(\ref{majorana})). The second equality in Eq.~(\ref{phasefactor}) holds because $d^\dagger d=0,1$ at zero temperature. In the fermionic operators we have defined above  the kinetic  energy  finally takes the simple form given in Eq.~(\ref{kinetic5}) while  the backscattering part of the Hamiltonian is
\begin{eqnarray}
H_{bs}&=&\frac{J_x}{\sqrt{2\pi a}}\left(\psi(0)+\psi^\dagger(0)\right){\cal P}\hat{S}_x\\&-&\frac{iJ_y}{\sqrt{2\pi a}}\left(\psi(0)-\psi^\dagger(0)\right){\cal P}\hat{S}_y.\nonumber
\end{eqnarray}
Using Eqs.~(\ref{phasefactor}) and (\ref{majorana}) together with the commutation relations for the operators $d$ and $d^\dagger$ we can show that ${\cal P}\hat{S}_x=-i\hat{S}_y$ and ${\cal P}\hat{S}_y=-i\hat{S}_x$. With these relations we recover Eq.~(\ref{bs7}).

\section{Green's functions}\label{green2}

Although this is rather standard material (see e.g. Ref.~\onlinecite{tsvelik}) we believe that it is useful to give a short derivation of the Green's functions Eq.~\ref{green}. The Fourier transforms of the  impurity Green's functions $G_a(\tau)=-\langle T_\tau a(\tau)a(0)\rangle$ and $G_b(\tau)=-\langle T_\tau b(\tau)b(0)\rangle$ can conveniently be found from the equations of motion. We will here only derive the correlation function for $a$, the correlator for $b$ can be found along the same lines. The Hamiltonian $H^A_{EK}=H_{Kin}+H_{bs}$ of our system is given in Eqs.~(\ref{kinetic5}) and (\ref{bs7}). The fields $\psi$ and $d=a+ib$ (see Eq.~(\ref{majorana})) obey standard fermionic anti-commutation relations. Introducing the Majorana components of the field $\psi$ through $z_1(x,\tau)= \left( \psi^\dagger(x,\tau)+\psi(x,\tau)\right)/\sqrt{2}$ and $z_2(x,\tau)= \left( \psi(x,\tau)-\psi^\dagger(x,\tau)\right)/(i\sqrt{2})$ we see already from the Hamiltonian in Eq.~(\ref{bs7})  that $a$ couples only to $z_2$.  We introduce the additional propagator $G_{z_2 a}(x,\tau)=-\langle T_\tau z_2(x,\tau) a(0)\rangle$. The equations of motion for the two coupled correlators  $G_a(\tau)$ and $ G_{z_2 a}(x,\tau)$ are
\begin{eqnarray}
\partial_\tau G_a(\tau)&=&-\delta(\tau)+i\frac{J_y}{\sqrt{2\pi a}}G_{z_2 a}(0,\tau),\\
\partial_\tau G_{z_2 a}(x,\tau)&=&+iv_F \partial_x G_{z_2 a}(x,\tau)-i\frac{J_y}{\sqrt{2\pi a}}\delta(x)G_a(\tau).\nonumber 
\end{eqnarray}  
To solve these equations it is best to go to Fourier space making use of the relations
\begin{eqnarray}
G_a(\tau)&=&\frac{1}{\beta}\sum_{\omega_n}{\rm e}^{-i\omega_n\tau}G_a(\omega_n),\nonumber \\
 G_{z_2 a}(x,\tau)&=&\frac{1}{\beta}\sum_{\omega_n}\int\frac{dp}{2\pi}{\rm e}^{-i\omega_n\tau+ipx}G_{z_2 a}(p,\omega_n),
\end{eqnarray}
where the sum is over the fermionic Matsubara frequencies $\omega_n=(2n+1)\pi/\beta$. To calculate the correlator $G_a(\tau)$ we only need to understand the local physics in $x=0$. The local equations of motion in Fourier space are 
\begin{eqnarray}
i\omega_nG_a(\omega_n)&=&1-i\frac{J_y}{\sqrt{2\pi a}}G_{z_2 a}(\omega_n),\label{b1}\\
G_{z_2 a}(\omega_n)&=&i\frac{J_y}{\sqrt{2\pi a}}G^{(0)}(\omega_n)G_a(\omega_n).\label{b2}
\end{eqnarray}
To alleviate the notation we have introduced  $G^{(0)}(\omega_n)=-i\,{\rm sgn}(\omega_n)/2v_F$. To find $G^{(0)}(\omega_n)$ we Fourier transform  the free electron propagator $G^{(0)}(p,\omega_n)=\left(i\omega_n-v_Fp\right)^{-1}$ with regard to $p$ and take the limit $x\rightarrow 0$. Furthermore we have defined $G_{z_2 a}(\omega_n)=G_{z_2 a}(x=0,\omega_n)$. 

Substituting Eq.~(\ref{b2}) into  Eq.~(\ref{b1}) we can solve for $G_a(\omega_n)$ and obtain
\begin{equation}
G_a(\omega_n)=\frac{1}{i\omega_n+i\Gamma_a sgn(\omega_n)}.
\end{equation}
After an analytic continuation $i\omega_n\rightarrow \omega+i\delta$ we recover Eq.~(\ref{green}).

\section{Mapping to the Kondo model in the small transmission limit}\label{t_small}

 In this Appendix first we want to fill in the gaps between Eq.~(\ref{effective}) and the Kondo Hamiltonian Eq.~(\ref{kondo}). 
Let us consider the tunneling part of the effective Hamiltonian. The first term takes an electron from the 2DEG and transfers it to the QD, the projection operator $P_0$ makes sure that the charge on the dot is $Q=0$ before the tunneling takes place. The second term takes an electron from the dot to the lead. In our truncated system this process is allowed only when the charge on the dot is $Q=1$. This restriction is implemented through the  operator $P_1$. The main goal of the following manipulations will be to show the equivalence of the Hamiltonian Eq.~(\ref{effective}) to a Kondo Hamiltonian. 

To explicitly account for the charge on the dot we make the replacement $|\Phi\rangle \rightarrow |\Phi\rangle |Q\rangle$. Here $|\Phi\rangle$ is any state of our system with charge $Q$ on the dot. The values of $Q$ are limited to $Q=0,1$ and the states $|Q\rangle=|0\rangle$ and $|Q\rangle=|1\rangle$ can be considered as the basis of a two-dimensional vector space.  However, the product $|\Phi\rangle |Q\rangle$ is no tensor product since the charge of the dot is of course not independent of the system's state. The state $|Q\rangle$ should rather be considered as an auxiliary label to $|\Phi\rangle$. In addition to introducing the label  $|Q\rangle$ we make the replacement
\begin{eqnarray}
\Psi^\dagger_{\sigma,1}\Psi_{\sigma,0}P_0&\rightarrow &\Psi^\dagger_{\sigma,1}\Psi_{\sigma,0}S^+,\\
\Psi^\dagger_{\sigma,0}\Psi_{\sigma,1}P_1&\rightarrow &\Psi^\dagger_{\sigma,0}\Psi_{\sigma,1}S^-\nonumber
\end{eqnarray}
in  Eq.~(\ref{effective}). Here $S^+$ and $S^-$ are pseudo-spin ladder operators acting only on the charge part $|Q\rangle$. 

Since $S^+|Q=1\rangle=0$ and $S^-|Q=0\rangle=0$ these operators ensure in the same way as the projection operators $P_0$ and $P_1$ that only transitions between states with $Q=0$ and $Q=1$ take place. In addition the charge $Q$ on the dot is adjusted whenever a tunneling process takes place since $S^+|0\rangle=|1\rangle$ and $S^-|1\rangle=|0\rangle$. We would like to emphasize again that only the combinations of pseudo-spin ladder operators and hopping operators introduced above are meaningful since $|Q\rangle$ and $|\Phi\rangle$ are not independent. 

To get rid of the remaining projection operators in Eq.~(\ref{effective}) we rewrite $(eUP_1)$ as $[eU(P_0+P_1)/2+eU(P_1-P_0)/2]$. We observe that 
\begin{eqnarray}
(P_1\pm P_0)|0\rangle &=&\pm |0\rangle,\\
(P_1\pm P_0)|1\rangle &=&+ |1\rangle.\nonumber  
\end{eqnarray}
This leads us to identify $(P_1-P_0)$ with $2S_z$ and $(P_1+P_0)$ with the identity operator on the space spanned by $|0\rangle$ and $|1\rangle$. Here we used that for the $z$-component $S_z$ of the pseudo-spin we have $S_z|1\rangle=|1\rangle/2$ and $S_z|0\rangle=-|0\rangle/2$. Gathering all terms we can rewrite the effective Hamiltonian Eq.~(\ref{effective}) as  
\begin{eqnarray}\label{effective2}
H_{eff}&=&\left(H^0_{kin}+H^1_{kin}\right)+eU(2S_z+1)/2\\
&+&\sum_{\sigma=1,2}\left(|t_\sigma| \Psi^\dagger_{\sigma,1}\Psi_{\sigma,0}S^++|t_\sigma| \Psi^\dagger_{\sigma,0}\Psi_{\sigma,1}S^-\right).
\end{eqnarray}
 We now introduce an additional pseudo-spin operator $s^{\pm}_{\sigma}(x)$ 
via
\begin{eqnarray}
s^{+}_{\sigma} &=& \Psi^\dagger_{\sigma,0}\Psi_{\sigma,1}\\ \nonumber
s^{-}_{\sigma} &=& \Psi^\dagger_{\sigma,1}\Psi_{\sigma,0},
\end{eqnarray}
where the matrices $\sigma^\pm=\sigma_x\pm i\sigma_y$ are standard combinations of Pauli matrices. It is important to note that these pseudo-spin operators again have nothing to do with the true spin of the electrons but are related to the location of an electron ($\alpha =0$ for an electron in the 2DEG, $\alpha =1$ for an electron in the QD). Introducing the definition of these pseudo-spins into  Eq.~(\ref{effective2}) finally leads us to  Eq.~(\ref{kondo}).

Finally, we want to show that in this limit 
the bosonization approach does not allow
us to precisely build a pseudo-spin operator describing the dot 
from the original 
tunnel Hamiltonian. 
For simplicity, we restrict the discussion to the case of spin-less 
fermions, i.e., we ignore the spin index $\sigma$.
The main problem we encounter is that
 at $x=0$ we have open boundaries, implying that
$\Psi_1(0)=\Psi_0(0)=0$. Introducing right (R) and left (L) movers as in Eq.
(\ref{boso}), this is equivalent to write, e.g. for the dot,
$\Psi_{1,R}(0)+\Psi_{1,L}(0)=0$. This has the effect to pin the (charge) fields
$\phi_{1}$ and $\phi_{0}$ at $x=0$: $\phi_{1}(0)=\phi_{0}(0)=
\sqrt{\pi}/2$. Therefore, this provides us
\begin{equation}
\Psi_{\alpha,p}(0)=\frac{\mp i}{\sqrt{2\pi a}}e^{i\sqrt{\pi}\theta_{\alpha}(0)}.
\end{equation}
From the form of the tunnel term (which can be rewritten either with
$\Psi_{1,R}$ or with $\Psi_{1,L}$) we would be tempted to explicitly 
build the pseudo-spin operator in the dot, as \cite{Larkin} 
\begin{eqnarray}
S^+ &=& \Psi^{\dagger}_{1,R/L}(0)\\ \nonumber
S^- &=& \Psi_{1,R/L}(0)\\ \nonumber
S_z &=& \Psi^{\dagger}_{1,R/L}(0)\Psi_{1,R/L}(0)-1/2.
\end{eqnarray}
However, due to the {\it open} boundary condition at $x=0$, the fermion 
operator $\Psi_{1,R/L}$ at $x=0$ now only depends on the superfluid phase 
$\theta_1$\cite{kane}. 
Then, $S^+$ would commute with $S_z$ and then $\vec{S}$ 
would not be a quantum spin object. 
The only way to proceed in order to recover the (correct) Kondo mapping is to introduce the extra label $|Q\rangle$.

\end{multicols}

\onecolumn
\begin{center}
\begin{tabular}{||c|c|c|c||}
\tableline
Nbr of channels& Reflection or transmission coefficient & Capacitance at N=1/2 & Ref. \\
\tableline

1 & $|r_1|\ll 1$, ($|r_2|\rightarrow 1$) & {\it const.} & Ref.~\onlinecite{m1}\\
\tableline
2 & $|r_1|=|r_2|\ll 1$ & $-\ln|N-1/2|$ & Ref.~\onlinecite{m1} \\
\tableline
2 & $|r_1|\neq |r_2|\ll 1$ & $-\ln (|r_2|-|r_1|)$ & Ref.~\onlinecite{karyn}, Secs. \ref{anisotropic}-\ref{renorm}\\
\tableline
1 & $|t_1|\ll 1$, ($|t_2|\rightarrow 0$) & {\it const.} & Ref.~\onlinecite{m2} \\
\tableline
2 & $|t_1|=|t_2|\ll 1$ & $-\ln|N-1/2|$ & Ref.~\onlinecite{m2} \\
\tableline
2 & $|t_1|\neq|t_2|\ll 1$ & $-\ln(|t_1|-|t_2|$) & Sec.~\ref{sec_t_small} \\
\tableline
\end{tabular}
\end{center}

\vskip 0.05cm
{\it Table 1: We have here listed all the 
results for the capacitance of the quantum dot at $N\rightarrow 1/2$ and
the respective references. The divergence occuring at  $N\rightarrow 1/2$ in the channel-isotropic {\it two}-channel limit is cut off by a channel anisotropy for both weak and strong reflection at the QPC.}

\end{document}